\documentclass[aps,pra,10pt,twocolumn,superscriptaddress]{revtex4-2}
\usepackage{amsmath,bm,mathdots}
\usepackage{xcolor}
\usepackage{graphicx,comment}
\usepackage{siunitx}

\usepackage[whole]{bxcjkjatype} 
\usepackage[bookmarks=false,colorlinks=true,urlcolor=blue,citecolor=blue,linkcolor=blue,breaklinks=true]{hyperref}
\input glyphtounicode
\graphicspath{{figs/}}

\usepackage{txfonts}

\begin{document}
\title{Dirac Electrons in AC-Magnetic Fields:\\ $\pi$-Landau Levels and Chiral Anomaly-Induced Homodyne Effect}
\author{Sota Kitamura}
\affiliation{Department of Applied Physics, The University of Tokyo, Hongo, Tokyo,
113-8656, Japan}
\author{Takashi Oka}
\affiliation{Institute for Solid State Physics, University of Tokyo, Kashiwa
277-8581, Japan}
\affiliation{Trans-scale Quantum Science Institute, University of Tokyo, Bunkyo-ku, Tokyo 113-0033, Japan}
\date{\today}
\begin{abstract}

Floquet engineering, which involves controlling systems through time-periodic driving, is a method for coherently manipulating quantum materials and realizing dynamical states with novel functionalities. 
Most research in solid-state systems has focused on the use of AC-\textit{electric} fields as the controlling drive. In this study, we investigate the effects of AC-\textit{magnetic} fields on two-dimensional (2D) Dirac electrons and report the emergence of new states and new transport phenomena. 
In a magnetic field that temporarily changes its direction, the 2D Dirac electrons form a new localized state with a flat band dispersion, dubbed as a $\pi$-Landau level. Its wave function is a superposition of the clockwise and counterclockwise cyclotron orbits with time-periodic amplitudes, resulting in a novel closed trajectory shaped like a figure eight.
Then, what would be the counterpart of the Hall effect in AC-magnetic fields? 
We find that a DC-current in the transverse direction, \textit{i.e.}
a homodyne Hall current, is generated when an additional AC-electric field is applied. 
In the case of Dirac electrons, several electronic states contribute to this phenomenon including the $\pi$-Landau level. 
However, when the chemical potential $\mu$ is near the Dirac point, 
the dominant contribution comes from the low-energy electrons and
we numerically find the homodyne Hall current to behave as $I_y=-\frac{e}{h}\mu$
per valley and spin. 
We explain this phenomenon through the high-frequency effective Floquet Hamiltonian 
which resembles the chiral Landau level Hamiltonian of three-dimensional Weyl Hamiltonian exhibiting chiral anomaly. 
We discuss the experimental feasibility and conclude that it is possible to realize this new exotic state
using techniques such as THz metamaterial enhancement of magnetic fields. 
\end{abstract}
\maketitle

\section{Introduction and Summary}
Electrons in static and homogeneous magnetic fields exhibit exotic transport properties, such as the quantum Hall effect. Classically, the dynamics of these electrons are described by a circular motion known as cyclotron motion, while in quantum systems, they are characterized by Landau orbits. When a static electric field is applied to a quantum Hall state, a current perpendicular to the field is induced, with their linear relation $j_x = \sigma_{xy} E_y$ defined by the Hall conductivity $\sigma_H = \frac{e^2}{h} \nu$. In the integer quantum Hall effect (IQHE), the factor $\nu$ is strictly an integer and is related to a topological index, the first Chern number, as shown by Thouless, Kohmoto, Nightingale, and den Nijs~\cite{Thouless1982}.

The quantum Hall effect has been extensively studied in Dirac electrons in both two-dimensional (2D)~\cite{CastroNeto2009,Novoselov2004,Novoselov2005,Zhang2005} and three-dimensional (3D)~\cite{Murakami_2007,Burkov2011,Armitage2018,Yan2017, Burkov2018} materials. Quantum anomalies play a crucial role in understanding the exotic transport phenomena observed in Dirac and Weyl fermions in various quantum materials. One such manifestation of anomaly is the chiral magnetic effect, which occurs in the presence of magnetic fields and results in an imbalance in the number of left- and right-handed fermions, leading to a charge current parallel to the applied field. The origin of these exotic phenomena can be traced back to the structure of the fermion spectrum in magnetic fields.

In static and homogeneous magnetic fields, the eigenstates of electrons form Landau levels, which are macroscopically degenerate in energy. Within the Landau levels of Weyl fermions, a special band known as the chiral Landau level appears as the zeroth Landau level~\cite{Nielsen1981}. This chiral Landau level carries nonzero current in the field direction, detectable in the presence of an imbalance, such as negative magnetoresistance and heat transport~\cite{Chernodub2022}. Furthermore, the zeroth Landau level is known for its robustness against disorder and serves as a platform for various strongly-correlated phenomena, including flatband ferromagnetism, the fractional quantum Hall effect, and magnetic catalysis~\cite{Gusynin1994}.

Studies on charged particles in magnetic fields have explored scenarios beyond homogeneous and static conditions. For example, when the magnetic field oscillates spatially with sign changes, a new conducting mode known as the snake state emerges~\cite{Mueller1992}. This mode propagates along a wiggling orbit and is strongly localized along the contours where the magnetic field vanishes. Snake states can be realized in materials such as carbon nanotubes under a perpendicular magnetic field, leading to peculiar anisotropic behavior in their magnetoresistance, especially in rolled-up geometries~\cite{Chang2014}.

Temporal oscillations of the magnetic field, \textit{i.e.}, AC-magnetic fields, can induce significant changes in electron motion. 
Experimentally, it is possible to generate strong AC-magnetic fields using 
metamaterial enhancement techniques~\cite{Liu2012,Iwaszczuk:12,Kozina2017,Bahk2017} and the created magnetic fields can be as strong as several Teslas, and the field 
oscillating in the THz regime~\cite{Mukai2014,Mukai_2016,Polley_2018,Qiu:18}. 
Previously, the AC-magnetic field-induced dynamics has been studied for non-relativistic electrons with quadratic dispersion~\cite{Oka2016} using Floquet theory, which describes quantum systems under periodic driving~\cite{Eckardt2017,Oka2019,Rudner2020}. 
In such conditions, electrons exhibit a snake-like wiggling motion due to sign flips of the curvature of their cyclotron motion following the magnetic field changes.

At specific ``magic" frequencies, a new feature akin to Landau quantization emerges where the electron trajectories form closed orbits resembling a figure of eight. In this regime, the effective mass of the electrons' center of mass coordinate diverges, rendering them immobile against static electric fields. When the quantum version of this problem is considered, these closed classical orbits result in macroscopically degenerate flat bands in the Floquet quasienergy spectrum. These Floquet Landau levels are characterized by a dissipationless frequency-converting Hall response.

\begin{figure*}[t]
\centering
\includegraphics[width=.9\linewidth]{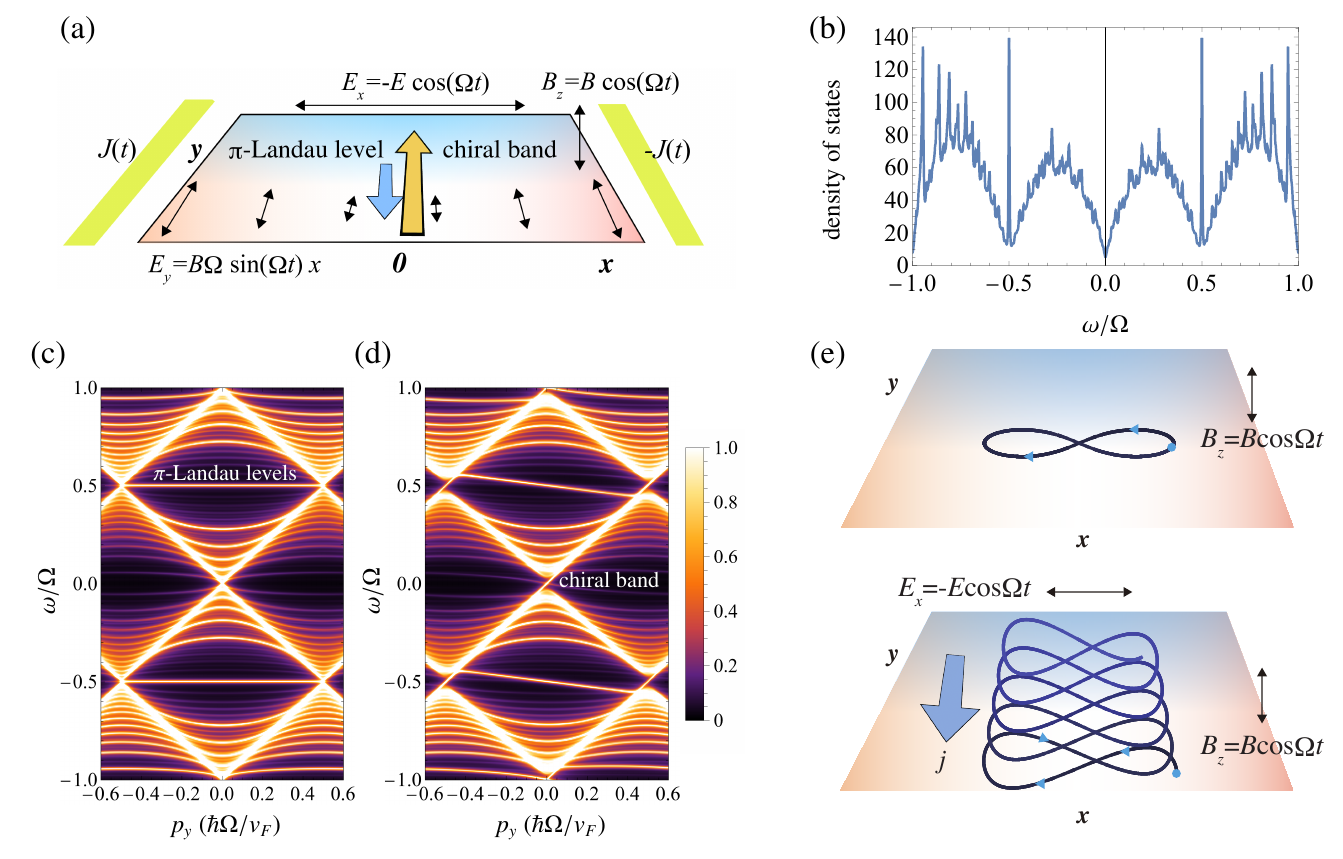}
\caption{(a) Schematic plot of the chiral band and $\pi$-Landau level that emerges at the center of the system. (b) Time average of the density of states. 
(c,d)
The quasienergy spectrum of the 2D Dirac fermions for
$B=2\hbar/el_0^{2}$, $E=0$ (c) and
$B=2\hbar/el_0^{2}$, $E=0.8\hbar\Omega/el_0$ (d).
The intensity is normalized by the peak value for the undriven case.
(e) Schematic image of the states in the $\pi$-Landau levels based on semiclassical trajectories.}
\label{fig:linear}
\end{figure*}

The concept of Floquet Landau levels and frequency-converting quantum Hall effect in AC-magnetic fields motivates us to study its interplay with the physics of quantum anomaly in Dirac and Weyl materials. 
In this paper, we study the energy spectrum and transport phenomena of 2D Dirac electrons and electrons on the honeycomb lattice under AC magnetic fields. We report rich physics closely connected to the chiral anomaly of three-dimensional (3D) Weyl fermions. 
In particular, we focus on a system described by the massless 2D Dirac Hamiltonian 
\begin{equation}
H=\tau v_F(\sigma_x\hat{\pi}_x+\sigma_y\hat{\pi}_y),
\label{eq:linearlized-model}
\end{equation}
and its lattice realization. 
Here, $\hat{\pi}_i=-i\hbar \partial_i+eA_i\;(i=x,y)$ denotes 
the kinetic momentum, $\tau=\pm 1$ is the chirality index, $v_F$ is the Fermi velocity, and $\sigma_{i}$ represents the Pauli matrices. 
The AC-magnetic field in $z$-direction is induced by the gauge field 
\begin{equation}
A_y=B\cos(\Omega t)x
\label{eq:Ay}
\end{equation}
which also adds a spatially varying electric field in the $y$-direction $-\dot{A}_y$. 
In addition to the $B$ field, we add an AC-electric field in the $x$-direction described by 
\begin{equation}
A_x=\frac{eE}{\Omega}\sin(\Omega t).
\end{equation}
The configuration of the AC fields is depicted in Fig.~\ref{fig:linear}(a). 
To realize the 2D Dirac electron in an AC magnetic field, one can place a single-layer graphene flake between the two wires [depicted as 
yellow lines with $\pm J(t)$] or construct a metamaterial structure on the surface of a 3D topological insulator that hosts surface 2D Dirac electrons. 
Then, the AC-magnetic field Eq.~(\ref{eq:Ay}) can be approximately generated as a radiation from the antiparallel current on two wires $J_y(x,t)=I_0 \cos(\Omega t) [\delta(x-L)-\delta(x+L)]$ at $|x|/L\ll 1$, as $B=\mu_0 I_0/\pi L$. 

An intriguing structure of the present system is evident 
in the numerically-computed energy spectrum $\overline{A}(\omega)$ [see Eq.~(\ref{eq:Abar})] of 2D Dirac fermion in an AC-magnetic field, shown in Fig.~\ref{fig:linear} (details are given later). 
The highlight features are the following. 
\begin{description}
    \item[$\pi$-Landau levels at $\omega=\pm\frac{\Omega}{2}$]
    When $B$ is nonzero but $E$ is zero, flat bands appear at $\omega=\pm\frac{\Omega}{2}$ that is surrounded by a series of arc-like bands that merges at $p_y=\pm\hbar\Omega/2v_F$ [Fig.~\ref{fig:linear}(c)]. 
    Each state is doubly degenerate (in terms of Floquet states). 
    When the AC-electric field $E$ is also switched on, the flat band at $\omega=\frac{\Omega}{2}$ tilts with a slope given by $eEv_F^2/\hbar\Omega^2$ turning it into a doubly degenerate chiral band [Fig.~\ref{fig:linear}(d)]. 
    \item[Chiral band at $\omega=0$] When both $B$ and $E$ are nonzero, a chiral band with slope $v_F$ is formed at zero energy ($\omega=0$). This state is nondegenerate.  
    We will show later that the energy dispersion obtained here is quite similar to that seen in the chiral magnetic effect of the 3D Weyl fermion in a static magnetic field. An interesting point in this regard is that the corresponding chiral anomaly here emerges for the vector U(1) symmetry, and the chiral fermions are \textit{both} right movers for two chiralities $\tau=\pm1$.
\end{description}

As we see in this paper, these spectral features can be captured by the Floquet effective Hamiltonian combined with a time-periodic unitary transformation. 
This also allows us to obtain analytically the approximate wave function of the $\pi$-Landau level. 
The $\pi$-Landau level wave function can be interpreted as a state that resonates between clockwise and counterclockwise cyclotron orbits and has a trajectory akin to the figure of eight, as schematically depicted in Fig.~\ref{fig:linear}(e). The orbital angular momentum carried by the $\pi$-Landau level turns out to show time-periodic oscillation synchronized with the AC magnetic field, consistent with the above picture.
This picture is also supported by a semiclassical analysis, where the closed orbit with the shape of a figure eight is obtained for an arbitrary driving frequency. 
We also show that the flat band is stable against perturbations, thanks to the dynamical chiral symmetry of the present system.

The existence of the chiral band at $\omega=0$ leads us to anticipate the transport analogous to the chiral magnetic effect. Namely, assuming an equilibrium distribution, the presence of the anomalous chiral band leads to the electric current expressed by
\begin{equation}
I_y=-\frac{e}{h}\mu
\label{eq:homodynecurre}
\end{equation}
with $\mu$ being the chemical potential measured from the charge neutral point. 
A homodyne effect refers to a generation of DC response arising from two AC inputs, in the present case $B$ and $E$, with the same frequencies $\omega$. 
Thus, a DC current described by Eq.~(\ref{eq:homodynecurre}) is a homodyne current. 
Since it is not linearly dependent on $B$ and $E$, in fact it is 
independent of them which we numerically confirm later, 
it is implied that this homodyne response has a purely nonperturbative mechanism.
Since the origin of this current can be traced back to the chiral Landau level at $\omega=0$ that has a deep relation with the 
chiral anomaly in 3D Weyl fermions, we refer to this effect as the 
\textit{chiral anomaly-induced homodyne effect} in this work.

The rest of this paper is organized as follows.
In Sec.~\ref{sec:preliminaries}, we give a brief review of the basic concepts such as properties of the chiral Landau level under the static magnetic field. In Sec.~\ref{sec:semiclassical}, we first discuss an intuitive understanding of the $\pi$-Landau level based on the semiclassical analysis of the wave packet, which results in the closed orbit solution with the shape of a figure eight. Then we show the energy spectrum of the 2D Dirac fermion and its lattice realization  in the presence of the AC-magnetic field, in Sec.~\ref{sec:quasienergy}. We then construct the Floquet effective Hamiltonian for the chiral band and the $\pi$-Landau levels using the series expansion technique, with which we reveal the chiral nature of the present system. 
In Sec.~\ref{sec:robustness},
we discuss the robustness of the flat band from the viewpoint of the dynamical symmetry, and its breakdown in a strong field. Finally, we discuss the DC Hall current against the AC electric field arising from the anomalous property in Sec.~\ref{sec:current}.
We give a conclusion in Sec.~\ref{sec:discussion}.

\section{Preliminaries}\label{sec:preliminaries}
Here, we give a brief review of several concepts that will be used in this work. 

\subsection{Chiral Landau level in static magnetic fields}
Here we review the physics of Landau levels in 2D Dirac and 3D Weyl fermions. 
\subsubsection{Two-dimensional case}\label{sec:preliminaries-2d}
For 2D case, the Hamiltonian in the presence of the static magnetic field $B$ is given by 
\begin{equation}
H_{\text{2D}}=\tau v_F[\sigma_{x}(-i\hbar\partial_{x})+\sigma_{y}(p_{y}+eBx)]\label{eq:static-2DDirac}
\end{equation}
in the Landau gauge. 
Let us define an annihilation operator of harmonic oscillator by 
\begin{equation}
\hat{a}=\frac{1}{\sqrt{2\hbar e|B|}}[-i\hbar\partial_{x}-is(p_{y}+eBx)]
\end{equation}
with $s=\text{sgn}B$,
which satisfies the canonical commutation relation $[\hat{a},\hat{a}^{\dagger}]=1$.
For $B>0$, the Hamiltonian in the matrix representation is written as 
\begin{equation}
H_{\text{2D}}=\tau v_F\sqrt{2\hbar e|B|}\begin{pmatrix}0 & \hat{a}\\
\hat{a}^{\dagger} & 0
\end{pmatrix}.\label{eq:2DDirac-static}
\end{equation}
Since $\hat{a}^{\dagger}\hat{a}+\sigma_{z}/2$ is a conserved quantity, 
we can obtain the eigenstate of the Hamiltonian $H_{\text{2D}}|v_{n,\pm}\rangle=E_{n,\pm}|v_{n,\pm}\rangle$ as
\begin{equation}
E_{n,\pm}=\pm\tau v_F\sqrt{2\hbar e|B|n},\quad|v_{ n,\pm}\rangle=\frac{1}{\sqrt{2}}\begin{pmatrix}\pm|n-1\rangle\\
|n\rangle
\end{pmatrix}\label{eq:static-2D-state-positive}
\end{equation}
with $n=1,2,\dots$, where $|n\rangle$ is the eigenstate of the number operator $\hat{a}^{\dagger}\hat{a}$ given by
\begin{equation}
|0\rangle=\left(\frac{e|B|}{\pi\hbar}\right)^{1/4}\int dxe^{-(p_{y}+eBx)^{2}/2\hbar e|B| }|x\rangle,\quad|n\rangle=\frac{(\hat{a}^{\dagger})^{n}}{\sqrt{n!}}|0\rangle.
\end{equation} 
These eigenstates satisfy the relation $\Gamma|v_{n,\pm}\rangle=-|v_{n,\mp}\rangle$ with $\Gamma=\sigma_{z}$ being the chiral operator, 
which is a consequence of the chiral symmetry $\{H,\Gamma\}=0$.

Along with the above eigenstates, there is a special eigenstate with zero energy 
\begin{equation}
E_0=0,\quad
|v_0\rangle=\begin{pmatrix}0\\
|0\rangle
\end{pmatrix}.\label{eq:static-2D-state-negative}
\end{equation}
This zero energy state is the simultaneous eigenstate of the chiral operator $\Gamma$ (with the eigenvalue $\Gamma=-1$) and has no partner state.
The case of $B<0$ can also be calculated straightforwardly,
where the eigenenergies are unchanged while the eigenstates are modified as
\begin{equation}
|v_0\rangle=\begin{pmatrix}|0\rangle\\
0
\end{pmatrix},\quad
|v_{n,\pm}\rangle=\frac{1}{\sqrt{2}}\begin{pmatrix}|n\rangle\\
\pm|n-1\rangle
\end{pmatrix}
\end{equation}
with $\Gamma|v_{0}\rangle=|v_{0}\rangle$, $\Gamma|v_{n,\pm}\rangle=|v_{n,\mp}\rangle$.

The eigenstates derived above carry orbital angular momentum and associated orbital magnetic moment as follows. The expectation value of the orbital angular momentum operator
\begin{align}
\hat{L}_{z}&=x(p_{y}+eBx)-yp_{x}\\
&=2x(p_y+eBx)+\frac{i[xy,H^{2}]}{2\hbar v_{F}^{2}}
\end{align}
can be calculated as
\begin{align}
\langle v_{n,\pm}|\hat{L}_{z}|v_{n,\pm}\rangle	&=s\hbar\langle v_{n,\pm}|(2\hat{a}^{\dagger}\hat{a}+1)|v_{n,\pm}\rangle=2ns\hbar,\label{eq:LL-Lz}\\
\langle v_{0}|\hat{L}_{z}|v_{0}\rangle	&=s\hbar,
\end{align}
where we have used $\langle v|\hat{a}|v\rangle=\langle v|\hat{a}^2|v\rangle=\langle v|[O,H^2]|v\rangle=0$.
All the states including the chiral eigenstate carry the mechanical angular momentum with the same sign according to the direction of the magnetic field $s=\text{sgn}B$.

On the other hand, the orbital magnetic moment $\hat{M}_{z}=(1/2)\bm{r}\times\bm{j}$ due to the electric current
\begin{align}
\hat{M}_{z}&=-\frac{1}{2}\tau ev_{F}(x\sigma_{y}-y\sigma_{x})\\
&=-\tau ev_{F}x\sigma_{y} -\frac{ie[xy,H]}{2\hbar}
\end{align}
has a sign factor reflecting whether the carrier is an electron or a hole.
The expectation value can be calculated as
\begin{align}
\langle v_{n,\pm}|\hat{M}_{z}|v_{n,\pm}\rangle	&=-i\tau \sqrt{\frac{e\hbar}{2|B|}}v_{F}\langle v_{n,\pm}|(\hat{a}-\hat{a}^{\dagger})\sigma_{y}|v_{n,\pm}\rangle\nonumber\\
	&=\mp\tau s\sqrt{\frac{ne\hbar}{2|B|}}v_{F},\label{eq:LL-Mz}\\
\langle v_{0}|\hat{M}_{z}|v_{0}\rangle	&=0.
\end{align}

A remarkable feature of the Landau level states is that the eigenenergy is independent of $p_{y}$, \textit{i.e.}, 
the energy eigenvalues have macroscopic degeneracy. The canonical momentum $p_{y}$ appears as the central position of the wave function as $x=-p_{y}/eB$. 
The degeneracy of the Landau level $D$ is given by $D=e|B|L^{2}/h$ with $L^{2}$ being the two-dimensional system size. Let us consider the response of the Landau levels against DC electric fields, $H_{E}=eEx$.
In terms of the harmonic oscillator, this perturbation is represented as 
\begin{equation}
H_{E}=\frac{E}{B}\left[
is\sqrt{\frac{\hbar e|B|}{2}}(\hat{a}-\hat{a}^\dagger)-p_{y}
\right].
\end{equation}
The first-order correction to the eigenenergy is obtained as 
$\langle H_{E}\rangle=-p_{y}E/B$
for all the Landau levels, with which the energy dispersion becomes tilted with the universal group velocity $-E/B$. 
Taking account of the degeneracy $D$, we obtain the DC Hall conductivity $\sigma_{yx}$ at the zero temperature as 
\begin{equation}
\sigma_{yx}=s\frac{e^{2}}{h}\left(n+\frac{1}{2}\right)
\end{equation}
when the Landau levels are occupied up to $n$-th one. 
Here we have assumed that the Hall current vanishes when the zeroth Landau level
is half-filled as it corresponds to the charge neutrality. 
Due to the Nielsen-Ninomiya theorem, we always have the Dirac fermion with $\tau=\pm1$ in a pairwise manner in crystalline systems, 
so that the above half-integer quantized value is doubled in the net current.

\subsubsection{Three-dimensional case}
Let us turn to the 3D Weyl fermion described by the Hamiltonian
\begin{equation}
H_{\text{3D}}=\tau v_F[\sigma_{x}(-i\hbar\partial_{x})+\sigma_{y}(p_{y}+eBx)+ \sigma_{z}p_{z}].\label{eq:static-3DWeyl}
\end{equation}
As the $p_{z}=0$ plane corresponds to the 2D Dirac case, 
let us take the 2D eigenstates $|v_{n,\pm}\rangle_{\text{2D}}$ as a basis set. 
Since $\Gamma=\sigma_{z}$ is the chiral operator in the 2D case, 
the matrix element in this basis can be calculated as 
\begin{gather}
H_{\text{3D}}|v_0\rangle_{\text{2D}}=-s\tau v_Fp_{z}|v_0\rangle_{\text{2D}},\\
H_{\text{3D}}|v_{n,\pm}\rangle_{\text{2D}}=\pm\tau v_F\sqrt{2\hbar e|B|n}|v_{n,\pm}\rangle_{\text{2D}}-s\tau v_Fp_{z}|v_{n,\mp}\rangle_{\text{2D}}.
\end{gather}
Namely, the zeroth Landau level remains to be the eigenstate of the 3D Hamiltonian,
$|v_0\rangle_{\text{3D}}=|v_0\rangle_{\text{2D}}$, 
but with the linear chiral dispersion 
\begin{equation}
E_{0}=-s\tau v_Fp_{z}.
\end{equation}
The remaining states constitute $2\times2$ block Hamiltonians, 
whose eigenvalues are given by
\begin{equation}
E_{n,\pm}=\pm \tau v_F\sqrt{2\hbar e|B|n+p_{z}^{2}}
\end{equation}
with $n=1,2,\dots$.

Due to the zeroth Landau level with the linear dispersion, 
the 3D Dirac fermion can carry nonzero current, 
while the higher Landau levels $n\neq0$ have vanishing current due to the even dispersion relation
$E_{n,\pm}(p_{z})=E_{n,\pm}(-p_{z})$. 
Assuming the vanishing current at the charge neutrality $\mu=0$, 
the electric current density is calculated as
\begin{align}
J_{z} & =s\tau ev_F \frac{D}{L^2} \int_{-\infty}^{\infty} \frac{dp_z}{2\pi\hbar} [f(-s\tau v_F p_z)-f(-s\tau v_F p_z)|_{\mu=0}]\nonumber\\
 & =\tau\frac{e^2}{h^2}\mu B
 \label{eq:CMEcurrent}
\end{align}
where $\mu$ is the chemical potential of the system, and 
$f(\varepsilon)=(1+e^{\beta(\varepsilon-\mu)})^{-1}$
is the Fermi-Dirac distribution function.
The result is independent of the Fermi velocity $v_F$, 
as well as the temperature $\beta^{-1}$.
This is known as the chiral magnetic effect.
This electric current due to the chiral dispersion cancels out when we add up the contribution of $\tau=\pm1$. 
A nonzero net current may appear when the chemical potential for the pair of Weyl fermions $\tau=\pm1$ are
different, 
which can be induced by applying DC electric field parallel to the magnetic field.

\subsection{Floquet formalism}
\subsubsection{Floquet quasi-energy spectrum}
To discuss the properties of the present model, we employ the Floquet theory for time-periodic problems~\cite{Eckardt2017,Oka2019,Rudner2020}.
The Floquet theorem assures that the solution of the time-dependent Schr\"odinger equation $i\hbar\partial_t|\psi(t)\rangle=H(t)|\psi(t)\rangle$ can be written in the form of the Floquet states, 
\begin{equation}
|\psi_{\alpha}(t)\rangle=|u_{\alpha}(t)\rangle e^{-i\varepsilon_{\alpha}t/\hbar},\quad|u_\alpha(t)\rangle=|u_\alpha(t+T)\rangle,
\label{eq:wf-floquet-form}
\end{equation}
with $T=2\pi/\Omega$ being the period of the Hamiltonian, $H(t)=H(t+T)$.

Since the periodic part of the wave function $|u_\alpha(t)\rangle$ can be expanded in the Fourier series, 
\begin{equation}
|u_\alpha(t)\rangle=\sum_{m=-\infty}^\infty|u_{\alpha,m}\rangle e^{-im\Omega t},
\end{equation}
we can rewrite the time-dependent Schr\"odinger equation as an eigenvalue problem in the extended Hilbert space (Sambe space) as
\begin{equation}
\begin{pmatrix}\ddots & \ddots & \ddots\\
\ddots & H_{0}+\hbar\Omega & H_{-1} & H_{-2}\\
\ddots & H_{+1} & H_{0} & H_{-1} & \ddots\\
 & H_{+2} & H_{+1} & H_{0}-\hbar\Omega & \ddots\\
 &  & \ddots & \ddots & \ddots
\end{pmatrix}\begin{pmatrix}\vdots\\
\vphantom{\vdots}|u_{\alpha,-1}\rangle\\
\vphantom{\vdots}|u_{\alpha,0}\rangle\\
\vphantom{\vdots}|u_{\alpha,+1}\rangle\\
\vdots
\end{pmatrix}=\varepsilon_\alpha
\begin{pmatrix}\vdots\\
\vphantom{\vdots}|u_{\alpha,-1}\rangle\\
\vphantom{\vdots}|u_{\alpha,0}\rangle\\
\vphantom{\vdots}|u_{\alpha,+1}\rangle\\
\vdots
\end{pmatrix},\label{eq:sambe}
\end{equation}
where $H_n$ is the $n$-th Fourier component of the time-periodic Hamiltonian $H(t)$.
Note that this eigenvalue problem has redundant solutions. Namely, $(\dots,|u_{\alpha,-1}\rangle,|u_{\alpha,0}\rangle,|u_{\alpha,+1}\rangle,\dots)^\text{T}$ and $(\dots,|u_{\alpha,m-1}\rangle,|u_{\alpha,m}\rangle,|u_{\alpha,m+1}\rangle,\dots)^\text{T}$ respectively have the eigenvalue of $\varepsilon_\alpha$ and $\varepsilon_\alpha+m\hbar \Omega$, but represent the same wave function $|\psi_\alpha(t)\rangle$.
A set of inequivalent solutions can be obtained by restricting the eigenvalue (called quasienergy) $\varepsilon_\alpha$ to the first Floquet Brillouin zone (BZ) $\varepsilon_\alpha\in (-\hbar\Omega/2,\hbar\Omega/2]$.

\subsubsection{Time-averaged energy spectrum}
The quasienergy corresponds to the energy defined for the coarse-grained dynamics averaged over the time period $T$.
Let us relate the quasienergy eigenvalue to the time average of the spectral function.
The spectral function can be calculated from the retarded Green function in the frequency domain. Its time average can be computed as
\begin{align}
\overline{A}(\omega) & =-\frac{1}{\pi}\text{Im}\int_{-\infty}^{\infty}dt\int_{0}^{T}\frac{d\tau}{T}\text{Tr}\,G_{0}^{R}\left(\tau+\frac{t}{2},\tau-\frac{t}{2}\right)e^{i\omega t},
\end{align}
where the retarded Green function is given by $[G_{0}^{R}(t,t^{\prime})]_{ij}=-i\langle\{c_{i}(t),c_{j}^{\dagger}(t^{\prime})\}\rangle\theta(t-t^{\prime})/\hbar$ for noninteracting systems, with $c_i$ being the annihilation operator of $i$-th electron.

Using the Floquet states that solves the eigenvalue problem in the Sambe space,  $G_0^R$ can be written as
\begin{equation}
[G_{0}^{R}(t,t^{\prime})]_{ij}=-\frac{i}{\hbar}\sum_{\alpha}\langle i|u_{\alpha}(t)\rangle\langle u_{\alpha}(t^{\prime})|j\rangle\theta(t-t^{\prime})e^{-i\varepsilon_{\alpha}(t-t^{\prime})/\hbar},
\end{equation}
where $|i\rangle=c_{i}^{\dagger}|0\rangle$ spans the one-particle Hilbert space, and $\alpha$ runs over the states in the first Floquet BZ. Accordingly, the time-averaged spectral function $\overline{A}(\omega)$ can be written with the Floquet states as
\begin{align}
\overline{A}(\omega) & =\sum_{\alpha m}\langle u_{\alpha,m}|u_{\alpha,m}\rangle\delta(\hbar\omega-\varepsilon_{\alpha}-m\hbar\Omega).
 \label{eq:Abar}
\end{align}

This expression tells us
how to unfold the quasienergy spectrum defined on $(-\hbar\Omega/2,\hbar\Omega/2]$ to the entire frequency domain.
We call this scheme to plot the quasienergy in the entire frequency domain the open Floquet BZ scheme.
The above expression can also be interpreted that the spectral weight is composed of the amplitude of the static component $\langle u_{\alpha,0}|u_{\alpha,0}\rangle$ for all the eigenvectors in the Sambe space, 
when the sum over the Fourier index $m$ is reinterpreted as that over the redundant solutions.

\subsubsection{Occupation function for a system coupled with a fermion bath}
The electrons' distribution of these bands is generically nonequilibrium and depends on how the system is coupled to dissipation sources. We here introduce an ideal fermionic reservoir, with which the system relaxes to a steady state with occupation 
\begin{equation}
f_\alpha=\sum_{m=-\infty}^\infty \langle u_{\alpha m}|u_{\alpha m}\rangle f(\varepsilon_{\alpha}+m\hbar\Omega)
\label{eq:falpha}
\end{equation}
for the $\alpha$-th Floquet state~\cite{Seetharam2015,Morimoto2016,Matsyshyn2023}.  
We give a brief derivation of this expression using the nonequilibrium Green function below. 

The self energy of the reservoir in terms of the nonequilibrium Green function is given by
\begin{align}
\Sigma^{R}(t,t^{\prime}) & =-\Sigma^{A}(t,t^{\prime})=-i\Gamma\delta(t-t^{\prime}),\\
\Sigma^{<}(t,t^{\prime}) & =i2\Gamma\int\frac{d\omega}{2\pi}f(\hbar\omega)e^{-i\omega(t-t^{\prime})},
\end{align}
where $f$ is the Fermi-Dirac distribution function (at the zero temperature). 
The retarded and advanced Green functions are then obtained as $G^R(t,t^\prime)=G^R_0(t,t^\prime)e^{-\Gamma(t-t^\prime)}=[G^A(t^\prime,t)]^\dagger$.
The lesser Green function $[G^{<}(t,t^{\prime})]_{ij}=i\langle c_{j}^{\dagger}(t^{\prime})c_{i}(t)\rangle/\hbar$ can be obtained via 
\begin{align}
G^{<}(t,t^{\prime})=
\int d\tau \int d\tau^\prime  G^{R}(t,\tau)\Sigma^{<}(\tau,\tau^{\prime})G^{A}(\tau^{\prime},t^{\prime}),
\end{align}
which can be computed as 
\begin{align}
[G^{<}(t,t^{\prime})]_{ij}=\frac{i}{\hbar}\sum_{\alpha\beta}f_{\alpha\beta}\langle i|u_{\alpha}(t)\rangle\langle u_{\beta}(t^{\prime})|j\rangle e^{-i\varepsilon_{\alpha}t/\hbar+i\varepsilon_{\beta}t^{\prime}/\hbar}\label{eq:lesser}
\end{align}
 with the occupation given by 
\begin{align}
f_{\alpha\beta} & =\sum_{mn}\int\frac{\Gamma\langle u_{\alpha m}|u_{\beta n}\rangle e^{i(\omega-\varepsilon_{\beta}/\hbar-n\Omega)t^{\prime}-i(\omega-\varepsilon_{\alpha}/\hbar-m\Omega)t}f(\hbar\omega)d\omega}{\pi(\omega-\varepsilon_{\alpha}/\hbar-m\Omega+i\Gamma)(\omega-\varepsilon_{\beta}/\hbar-n\Omega-i\Gamma)}\nonumber\\
 & \to\delta_{\alpha\beta}\sum_{m}\langle u_{\alpha m}|u_{\alpha m}\rangle f(\varepsilon_{\alpha}+m\hbar\Omega)\quad(\Gamma\to+0),
 \label{eq:occupation}
\end{align}
which is nothing but Eq.~(\ref{eq:falpha}). 
Here, we have assumed that the Floquet states are nondegenerate. 
In the degenerate case, $f_{\alpha\beta}$ for $\varepsilon_\alpha=\varepsilon_\beta$ is given by $\sum_{m} \langle u_{\alpha m}|u_{\beta m}\rangle f(\varepsilon_{\alpha}+m\hbar\Omega)$ in $\Gamma\to+0$, which is a Hermitian block matrix. Namely, the same expression as the nondegenerate case can be used for the linear combination diagonalizing the block matrix.

\section{Semiclassical analysis of the $\pi$-Landau level}\label{sec:semiclassical}
\begin{figure}[thb]
\centering
\includegraphics[width=\linewidth]{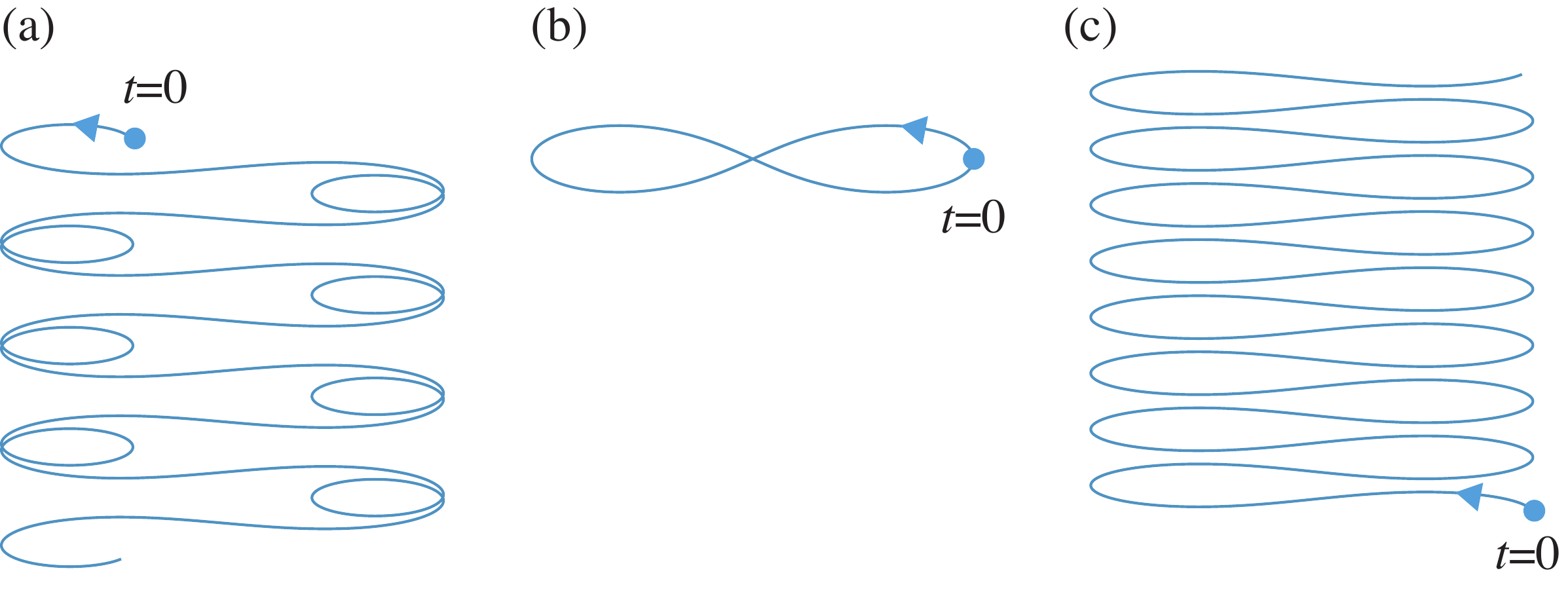}
\caption{Typical trajectories of the semiclassical wave packet under AC magnetic field Eqs.~(\ref{eq:classical1}) and (\ref{eq:classical2}).
The initial value of the momentum is taken to be (a) $|\bm{p}|=0.2p_0$, (b) $|\bm{p}|=0.42p_0$, and (c) $|\bm{p}|=0.5p_0$ with $p_0=eBv_F/\Omega$. The direction of the momentum at $t=0$ is taken to be $y$ direction ($p_y(0)=|\bm{p}|$).}
\label{fig:trajectory}
\end{figure}

Before starting the quantum analysis of the present problem, here we consider a semiclassical treatment assuming a linear dispersion relation $E=v_F|\bm{p}|$ with mechanical momentum $\bm{p}=\hbar\bm{k}+e\bm{A}$, to get an intuition for the emergence of the $\pi$-Landau level. The center-of-mass position and momentum of a two-dimensional electron wave packet under the AC magnetic field $B_z(t)=B\cos(\Omega t)$ are given by 
\begin{gather}
\dot{\bm{r}} =\frac{\partial(v_{F}|\bm{p}|)}{\partial\bm{p}}=v_{F}\frac{\bm{p}}{|\bm{p}|},
\label{eq:classical1}\\
\dot{\bm{p}} =-e\dot{\bm{r}}\times\bm{B}(t)=-e(\dot{y},-\dot{x})B_z(t),\label{eq:classical2}
\end{gather}
where we have neglected the contribution of the induced electric field
$\bm{E}=-\partial_{t}\bm{A}$ on the second line. We note that this approximation is adopted only in this section, and the influence of the induced electric field is fully taken into account in the quantum analysis. By substituting the first line into second, we obtain
\begin{equation}
\dot{p}_{x}+i\dot{p}_{y}=i\frac{ev_{F}}{|\bm{p}|}(p_{x}+ip_{y})B_z(t),\quad\frac{d|\bm{p}|}{dt}=0.
\end{equation}
This can be solved as
\begin{align}
p_{x}(t)+ip_{y}(t)&=|\bm{p}|e^{i\theta}\exp\left[i\frac{ev_{F}}{|\bm{p}|}\int_0^t dt^\prime B_z(t^\prime)\right]\\
&=|\bm{p}|e^{i\theta}\exp\left[i\frac{ev_{F}B}{|\bm{p}|\Omega}\sin(\Omega t)\right]
\end{align}
with $\theta=\tan^{-1}[p_{y}(0)/p_{x}(0)]$. Then, the center-of-mass position
of the wave packet can be obtained by integrating $\dot{x}+i\dot{y}=(v_{F}/|\bm{p}|)(p_{x}+ip_{y})$. We show typical trajectories in Fig.~\ref{fig:trajectory}.

The (signed) curvature of the trajectory $\kappa$ as a function of $t$ is given by $\kappa=(\dot{x}\ddot{y}-\dot{y}\ddot{x})/(\dot{x}^{2}+\dot{y}^{2})^{3/2}$, which can be caluclated as $\kappa=eB\cos(\Omega t)/|\bm{p}|$. Due to the sign change by the cosine factor, the trajectory of the wave packet center takes the snake-like shape as can be seen in Fig.~\ref{fig:trajectory}. While the trajectory is not closed in general, the trajectory becomes a closed orbit in the shape of a figure eight, when the amplitude of the phase factor $ev_FB/|\bm{p}|\Omega$ takes a special value. A similar solution is found in Ref.~\cite{Oka2016} for the case of a parabolic dispersion $E=|\bm{p}|^2/2m_e$, where the amplitude of the phase factor $ev_FB/|\bm{p}|\Omega$ is replaced by $eB/m_e\Omega$.


The condition for the closed orbit can be obtained analytically. 
Using the Jacobi-Anger identity
\begin{equation}
e^{iz\sin\theta}=\sum_{m=-\infty}^\infty J_m(z)e^{im\theta}
\end{equation}
with $J_n$ being the $m$th Bessel function, 
one can write $x+iy$ in a series form.
In particular, at the stroboscopic times $t=nT$ with integer $n$, the position is given by the $m=0$ term as
\begin{equation}
    x(nT)+iy(nT)=v_{F}e^{i\theta}J_{0}\left(\frac{ev_{F}B}{|\bm{p}|\Omega}\right)nT.
\end{equation}
This result indicates that the trajectory takes a closed orbit when the value of the Bessel function vanishes, which is satisfied, e.g., for $|\bm{p}|\simeq0.42eBv_{F}/\Omega$. This is in a sharp contrast to the parabolic case, where the fine-tuning of the driving frequency is necessary since the argument of the Bessel function $eB/m_e\Omega$ is independent of momentum.

The interval of the Landau levels in a static magnetic field can be obtained by the Sommerfeld quantization rule, as $h/T_c$ with $T_c$ being the period of the semiclassical motion. If we naively extend this argument to this time-dependent problem, the energy interval is predicted as $\hbar\Omega$, which indeed coincides with that of the $\pi$-Landau level shown in Fig.~\ref{fig:linear}(c). 
While the robustness of the presence of the closed orbit against the quantum effect and the neglected electric field is not clear from this analysis, in the later section we show that the $\pi$-Landau level wave function shares the intuitive picture of the figure-of-eight shaped cyclotron orbit.

\section{Floquet quasi-energy spectrum}\label{sec:quasienergy}
\subsection{Numerical computation of the energy spectrum}\label{sec:numerics}
In this section, we introduce the Hamiltonian of the present study, and show the numerical result for the quasienergy spectrum based on the formalism introduced in the previous section.

As we have introduced in Introduction, we consider the 2D Dirac fermion coupled to the AC-electromagnetic field
\begin{equation}
H(t)=\tau v_F\left[\sigma_x\left(-i\hbar\partial_x+\frac{eE}{\Omega}\sin\Omega t\right)+\sigma_y(p_y+eBx\cos\Omega t)\right],
\label{eq:linearlized-model2}
\end{equation}
where $v_F$ is the Fermi velocity, and $\tau$ is the chirality index.
The canonical momentum along $y$ is a good quantum number, and is denoted as $p_y$ here.
In this model, we have two characteristic length scales, $l_B=\sqrt{\hbar/eB}$ and $l_0=v_F T$. We take $l_0$ as the unit of length.

We use the matrix representation using the harmonic oscillator basis, given by $|n\rangle=(\hat{b}^\dagger)^n|0\rangle/\sqrt{n!}$ with
\begin{equation}
\hat{b}=\frac{1}{\sqrt{\hbar eB}}\left(-i\hbar\partial_{x}-i\frac{eB}{2}x\right).\label{eq:ho}
\end{equation}
See numerical details for Appendix~\ref{sec:appendix-numerics}.
We compute the Floquet state, i.e., the eigenstate of the Hamiltonian in the Sambe space [Eq.~(\ref{eq:sambe})], and plot the quasienergy spectrum using the open Floquet BZ scheme introduced above [\textit{i.e.}, calculate the time-averaged spectral function $\overline{A}(\omega)$ given by Eq.~(\ref{eq:Abar})]. 

Figures~\ref{fig:linear}(c),(d) depict the Floquet quasi-energy spectrum for several values of the AC-magnetic field $B$ and the AC-electric field $E$. 
The spectrum with no external fields [Fig.~\ref{fig:linear}(c)] has a Dirac point at $\varepsilon=0, p_y=0$ with continuous excitation.
With an applied AC-magnetic field $B=2\hbar/el_0^2$, as shown in Fig.~\ref{fig:linear}(c), the spectrum becomes discrete while keeping the dispersion around the Dirac point gapless. Here a remarkable point is the flat bands pinned at $\omega=\pm\Omega/2$, which we call $\pi$-Landau levels here.
The emergence of the $\pi$-Landau levels at the boundary of the Floquet Brillouin zone is rather surprising. This is because we expect absence of states at the zone boundary due to the level repulsion between the original band and a Floquet replica band that leads to a gap opening. In fact, this is how the anomalous Floquet topological insulator is realized where $\pi$-edge states emerges within the gap~\cite{FrederikAFI}.
In the present case, the electron and hole bands are hybridized by the field and form a flat band that is pinned at the Floquet BZ boundary. 

By further applying an AC-electric field $E=0.8\hbar\Omega/el_0$,
we obtain the energy spectrum shown in Fig.~\ref{fig:linear}(d).
The $\pi$-Landau levels now have a slope proportional to the field amplitude $E$, which implies the emergence of the DC Hall current.
Another remarkable feature here is the formation of the chiral band at zero energy. The slope of the chiral band is $v_F$, and independent of the field amplitude. An interesting point is that the slope is also independent of the chirality index $\tau=\pm1$, which apparently violates the Nielsen-Ninomiya theorem in one-dimensional static systems.

In order to further understand the nature of the 2D Dirac electrons in AC-magnetic fields, we  consider its realization on the honeycomb lattice tight-binding model described by the Hamiltonian
\begin{align}
H(t) & =t_{0}\sum_{ij}^{\text{N.N.}}e^{-i\bm{A}_{i}(t)\cdot(\bm{R}_{i}-\bm{R}_{j})}c_{iA}^{\dagger}c_{jB}+\text{H.c.}\label{eq:lattice-model}
\end{align}
Here, $t_0$ is the nearest-neighbor hopping amplitude, and $c_{jX}$ denotes the annihilation operator of the electron at the site $j$ on the sublattice $X\in\{A,B\}$. 
$A$ sites are located at $\bm{R}_i=(3n/2,\sqrt{3}(n/2+m))$ with $n,m\in\mathbb{Z}$, and 
the site summation is taken for the nearest-neighbor bonds, $\bm{R}_{i}-\bm{R}_{j}=(\cos2\pi l/3,\sin2\pi l/3)$ with $l=0,1,2$. 
In the lattice model, we introduce the external fields by the Peierls substitution with the site-dependent vector potential $\bm{A}_i(t)$, which is dimensionless in the present notation.
The fields are explicitly given by 
\begin{equation}
\bm{A}_{i}(t)=\left(\frac{eEa}{\hbar\Omega}\sin(\Omega t),\frac{eBa^{2}}{\hbar}x_{i}\cos(\Omega t)\right)
\end{equation}
where $a$ is the length of the nearest-neighbor bond ($\sqrt{3}a$ is the lattice constant), 
and $x_i=\bm{R}_i\cdot\bm{e}_x$ is the dimensionless coordinate along $x$. 
The low-energy limit of this lattice Hamiltonian coincides with Eq.~(\ref{eq:linearlized-model2}) via the relation $v_F =3t_0a/2\hbar$, which can also be written as $l_0/a=3\pi t_0/\hbar\Omega$.

\begin{figure*}[thb]
\centering
\includegraphics[width=\linewidth]{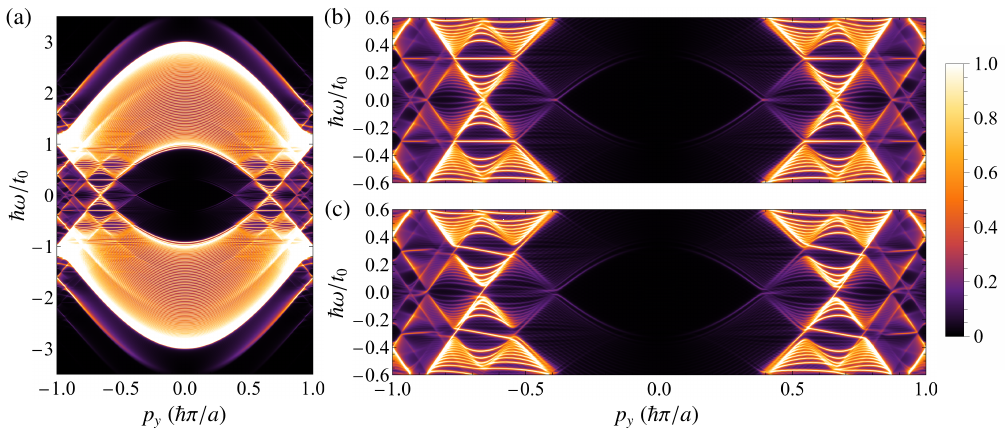}
\caption{Energy spectrum of the honeycomb lattice model driven by a time-periodic magnetic field, plotted as a function of the canonical momentum $p_y$ and the energy $\hbar\omega$. 
The spectral function is averaged over a driving period, and the intensity is normalized by the peak value for the undriven case.
(a) $B=2\hbar/el_0^{2}$.
(b) magnified view of (a).
(c) $B=2\hbar/el_0^{2}$ but with an additional AC electric field $E=0.8\hbar\Omega/el_0$.
}
\label{fig:spectrum}
\end{figure*}

We numerically calculate the Floquet quasienergy spectrum, as shown in Fig.~\ref{fig:spectrum}.
Here we choose the driving frequency as $\hbar\Omega=0.6t_0$, and 
the amplitude of the applied magnetic field as $B=2\hbar/el_0^2=0.008\hbar/ea^{2}$.
We adopt the open boundary condition, with which we have zig-zag edges along $y$ direction.
The number of sites along $x$ direction is set to $200$ [$100$ unitcells along $(3/2,\sqrt{3}/2)$], where the corresponding sample width is much larger than the magnetic length $l_B=\sqrt{\hbar/eB}=11a$.
Note that the electric field $E_y\propto x_i$ at the edges becomes larger as we increase the sample width, so that the agreement between the continuous and lattice model is expected only for finite-width cases.

We show the quasienergy spectrum under the AC-magnetic field in Fig.~\ref{fig:spectrum}(a), with a magnified view of the low-energy part in Fig.~\ref{fig:spectrum}(b). The spectrum under an additional AC electric field $E=0.8\hbar\Omega/el_0=0.05\hbar\Omega/ea$ is also shown in Fig.~\ref{fig:spectrum}(c).
We can see that the spectral features in the continuous Hamiltonian, \textit{i.e.}, the emergence of the $\pi$-Landau levels and the zero-energy chiral bands is well reproduced.
Indeed, the slope of the chiral modes is the same for the two valleys, as we have mentioned. This spectral structure is allowed as the present system is periodically-driven, where the spectral weight is not necessarily constant and allowed to vanish. 

Hereafter, we set $\hbar=e=1$. We also set $v_F=1$, while we keep the energy unit $\Omega$ as it is convenient for keeping track of the order of the perturbative expansion.

\subsection{Effective Hamiltonian for the chiral band at $\varepsilon=0$}\label{sec:quasienergy-2}
Let us first explore the origin of the chiral fermion at the zero quasienergy,
which emerges when an additional AC electric field $E$ is applied [Fig.~\ref{fig:linear}(d), Fig.~\ref{fig:spectrum}(c)]. 
To this end, we here derive an effective static Hamiltonian and show that the chiral fermion is related to the chiral Landau level of 3D Weyl fermion. 
A standard method to obtain a perturbative expression for the effective Hamiltonian is known as the high-frequency expansion~\cite{PhysRevB.84.235108,Eckardt2015,Bukov2015,Mikami2016}, 
which can be obtained by applying the van Vleck quasi-degenerated perturbation theory to Eq.~(\ref{eq:sambe}), 
with regarding $\Omega$'s in the diagonal entities as the unperturbed Hamiltonian. The Floquet effective Hamiltonian is given by~\cite{PhysRevB.84.235108}
\begin{equation}
H_\text{F}=H_0+\sum_{m\neq0} \frac{[H_{-m},H_m]}{2m\Omega}+O(\Omega^{-2}),\label{eq:hfe}
\end{equation}
which has the dimension same as the original time-dependent Hamiltonian, and does not contain any redundant solutions.
This gives a good approximation when the matrix elements of $H_m$ are sufficiently smaller than $\hbar\Omega$ between the eigenstates of $H_0$ we are interested in.

As depicted in Fig.~\ref{fig:linear}(c) and Fig.~\ref{fig:spectrum}(b), when the additional electric field is absent ($B\ne 0,\;E=0$), the
high-frequency expansion of Eq.~(\ref{eq:linearlized-model2})
leads to \textit{no} significant modification. The first-order correction $[H_{-1},H_{+1}]/\Omega$ vanish identically,
while the time average (zeroth-order term $H_0$) coincides with the undriven Hamiltonian. 
Thus the spectrum around $\omega=0$
remains linear in the numerically-obtained spectrum.

A nontrivial correction emerges when the AC electric field is switched on ($B\ne 0,\;E\ne 0$).
The AC electric field term $H_E(t)$ in terms of the linearlized model (\ref{eq:linearlized-model2}) is written as
\begin{equation}
H_E(t)= \tau\sigma_x \frac{E}{\Omega}\sin(\Omega t)\label{eq:electric-field},
\end{equation}
with which the high-frequency expansion (\ref{eq:hfe}) up to $1/\Omega$ correction results in
\begin{equation}
H_{\text{F}}=\tau \left[\sigma_{x}(-i\partial_{x})+\sigma_{y}p_{y} +\sigma_{z}\tau\frac{BE}{\Omega^2}x\right]
\label{eq:HFchiral}
\end{equation}
This Hamiltonian can be seen as a Hamiltonian of a 3D Weyl fermion projected to $p_z=0$ under an applied static magnetic field along $y$-axis [See Eq.~(\ref{eq:static-3DWeyl})]. An important point here is that the field strength $B_{\text{eff}}$ is given by 
\begin{equation}
B_{\text{eff}}=-\tau \frac{BE}{\Omega^2},
\end{equation} and depends on the chirality of the Weyl fermion $\tau$, as opposed to the usual case.
The Floquet spectrum around $\omega=0$ is obtained as the eigenvalues of Eq.~(\ref{eq:HFchiral}) and become
\begin{equation}
\varepsilon_0=p_{y},\qquad \varepsilon_{n,\pm}=\pm \sqrt{|B_{\text{eff}}|n+p_y^2}\quad (n=1,2,\ldots),
\end{equation}  
which is similar to the Landau levels in 3D Weyl fermions.  

There are a few comments to be noted. 
First, although the bands near $\omega=0$ are analogous to the Landau levels of 3D Weyl fermions, 
their degeneracy is different. In the case of the bands of Eq.~(\ref{eq:HFchiral}) they are non-degenerate due to the restriction $p_z=0$, 
which is in contrast to the degeneracy of $\mathcal{N}=eBL^2/h$ for the Landau level degeneracy in 3D Weyl fermions.   
Second, the slope of the chiral band $\varepsilon_0$ is common for both chiralities $\tau=\pm 1$.  
Thus, in the case of the honeycomb lattice, the $K$ and $K'$-points show a similar chiral band structure
as we can see in Fig.~\ref{fig:spectrum}(c). 
The direction of the chiral bands can be changed by flipping the sign of a product $BE$.

\subsection{Effective Hamiltonian for the $\pi$-Landau level at $\varepsilon=\pm \Omega/2$}\label{sec:quasienergy-3}
Next, we investigate the origin of the flat bands at the Floquet Brillouin zone boundary $\omega=\pm\Omega/2$.
As we have seen in the previous subsection, the flat band states 
are not described by the 
effective Floquet Hamiltonian within the high-frequency expansion [see Eq.~(\ref{eq:HFchiral})]. 
This is because the high-frequency expansion describes the physics 
of small quasi-energy, and not the states with quasi-energy of the order of $\Omega$. 
The kinetic energy operator $-i\partial_x$ takes a huge expectation value even when the field amplitude $B$ is small, which leads to the failure of the expansion.
This implies that we need to subtract contribution divergent in the high-frequency limit $\Omega\to\infty$ before performing the high-frequency expansion.

Indeed, we can carry out this subtraction using a time-periodic unitary transformation $U(t)=U(t+T)$ given by
\begin{equation}
U(t)=e^{i(\Omega/2)x\cos\theta}e^{-i\sigma_{z}\theta/2}e^{-i\tau(\Omega/2)(1+\sigma_{x})t},
\label{eq:rotating-transformation}
\end{equation}
where $\theta$ satisfies $p_y=(\Omega/2)\sin\theta$.
The transformed Hamiltonian for $E=0$ is obtained as
\begin{align}
H_{\text{rot}}(t) & =U^{\dagger}(t)(H(t)-i\partial_{t})U(t)\nonumber\\
 & =-\tau\frac{\Omega}{2}+\tau \cos\theta\left[\sigma_{x}(-i\partial_{x})+\sigma_{y}\frac{B}{2}x\right]\nonumber \\
 & -\tau\sigma_{x}^{+} \sin\theta e^{i\tau\Omega t}(-i\partial_{x})+\text{H.c.}\nonumber \\
 & +\tau (\sigma_{x}\sin\theta e^{i\Omega t}+\sigma_{x}^{+}\cos\theta e^{i\tau2\Omega t})\frac{B}{2}x+\text{H.c.},\label{eq:rotating-frame}
\end{align}
where $\sigma_{x}^{+}=(\sigma_{y}+i\sigma_{z})/2$.
In particular, 
the time average of the Hamiltonian $\overline{H_{\text{rot}}(t)}$ reads
\begin{equation}
\overline{H_{\text{rot}}(t)} =-\tau\frac{\Omega}{2}+\tau \cos\theta\left[\sigma_{x}(-i\partial_{x})+\sigma_{y}\frac{B}{2}x\right],
\label{eq:piLandau}
\end{equation} 
which is the leading-order term of the high-frequency expansion.

The unitary transformation introduced above can be decomposed into two steps, $U(t)=V_1V_2(t)$ with $V_1=e^{i(\Omega/2)x\cos\theta}e^{-i\sigma_{z}\theta/2}$ and $V_2(t)=e^{-i\tau(\Omega/2)(1+\sigma_{x})t}$.
At the first step $V_1$, we extract the $O(\Omega)$ contribution from the kinetic term; In the absence of the external field, the quasienergy eigenstate with
$\varepsilon_\alpha=\Omega/2$ is realized when $\langle-i \partial_{x}\rangle^{2}+p_{y}^{2}=(\Omega/2)^{2}$.
This implies that it is convenient to shift the origin of the momentum $-i\partial_{x}$
by $(\Omega/2)\cos\theta=\pm[(\Omega/2)^{2}-p_{y}^{2}]^{1/2}$.
While there are two choices of such $\theta$, we assume that the hybridization
between two plane waves is not relevant even when the external field
is switched on. 
Then the transformed Hamiltonian is expressed as
\begin{align}
V_1^{\dagger}H(t)V_1 & =\tau\frac{\Omega}{2}\sigma_{x}+\tau (\sigma_{x}\cos\theta-\sigma_{y}\sin\theta)(-i\partial_{x})\nonumber \\
 & +\tau (\sigma_{x}\sin\theta+\sigma_{y}\cos\theta)Bx\cos\Omega t,
 \label{eq:VHV}
\end{align}
where we have used $p_{y}=(\Omega/2)\sin\theta$. 
Now the kinetic term represents the residual fluctuation,
so that only the first term of Eq.~(\ref{eq:VHV}) is expected to have a divergent contribution of $O(\Omega)$ in the weak-field case.
Then, at the second step $V_2(t)$, we remove the first term by moving to the co-rotating frame, with which we obtain Eq.~(\ref{eq:rotating-frame}).

The first term of $\overline{H_{\text{rot}}(t)}$ gives the energy shift to the zone edge of the Floquet Brillouin zone. 
The remaining part is equivalent to the Hamiltonian for the static Landau levels at $p_y=0$ with a renormalization factor $\tau \cos\theta$ [See Eq.~(\ref{eq:static-2DDirac})].
Thus, we regard $\overline{H_{\text{rot}}(t)}$ as the effective Hamiltonian for the $\pi$-Landau levels that appeared in Figs.~\ref{fig:linear}(c), and \ref{fig:spectrum}.

The time-averaged Hamiltonian $\overline{H_{\text{rot}}(t)}$ can be diagonalized in
terms of the harmonic oscillator Eq.~(\ref{eq:ho}),
as in the Dirac electrons in a static magnetic field. The eigenstates
are given by 
\begin{equation}
|v_{0}\rangle=\begin{pmatrix}0\\
|0\rangle
\end{pmatrix}
,\quad|v_{n,\pm}\rangle=\frac{1}{\sqrt{2}}\begin{pmatrix}\pm|n-1\rangle
\\
|n\rangle
\end{pmatrix}
\label{eq:solution}
\end{equation}
with $n=1,2,\dots$, where the ground state of the harmonic oscillator $|0\rangle$ is given by 
$\langle x|0\rangle=(B/2\pi)^{1/4}\exp(-Bx^{2}/4)$. The eigenenergies
of these states are given by 
\begin{equation}
\varepsilon_{0}=-\tau\frac{\Omega}{2},\quad\varepsilon_{n,\pm}=-\tau\frac{\Omega}{2}\pm\tau \sqrt{B n}\cos\theta.
\label{eq:piLL-dispersion}
\end{equation}
The zeroth Landau level indeed gives a flat band, while the remaining
Landau levels form elliptic bands via 
\begin{equation}
\cos\theta=\pm\sqrt{1-\left(\frac{2p_{y}}{\Omega}\right)^{2}}.
\end{equation}
Since there are two choices for $\theta$, there are two Landau level
solutions for given $p_{y}$. This is consistent with the fact that
the numerically-obtained flat and elliptic bands are all doubly-degenerate. 
With these results, the time-averaged density of states for the Landau levels is obtained as 
\begin{align}
D(\varepsilon) & =D_{0}\left(\varepsilon+\frac{\Omega}{2}\right)+D_{0}\left(\varepsilon-\frac{\Omega}{2}\right),\\
D_{0}(\varepsilon) & =\frac{\Omega L_{y}}{2\pi}\left[\delta(\varepsilon)+\sum_{n=1}^{\infty}\frac{|\varepsilon|}{\sqrt{Bn}\sqrt{Bn-\varepsilon^{2}}}\right]
\end{align}
with $L_{y}$ being the system size along $y$ direction. Along with
the peak structure due to the flat band, elliptic bands also leads
to van Hove singularities with $\sim\varepsilon^{-1/2}$, which is consistent with the numerically-obtained result shown in Fig.~\ref{fig:linear}(b).

\begin{figure}[t]
\centering
\includegraphics[width=\linewidth]{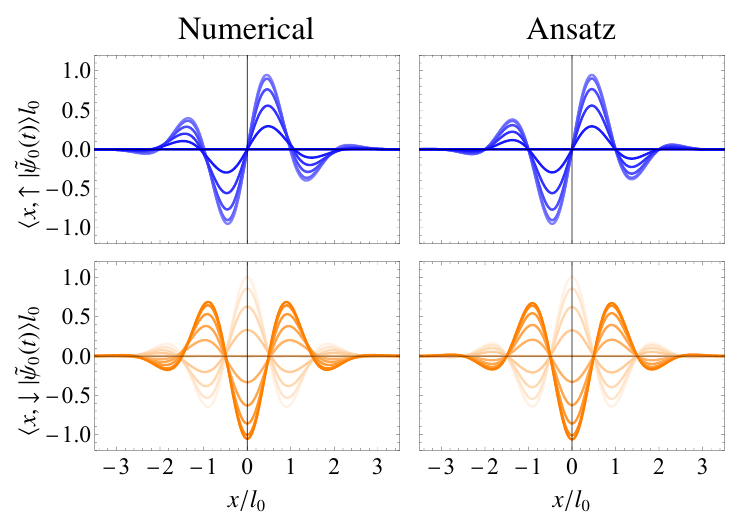}
\caption{The wave function of the flat band state at the boundary of the Floquet Brillouin zone for $B=2\hbar/el_0^{2}, p_y=0$, compared between the numerical result and the ansatz obtained in the rotating frame. We show snapshots at $t=0.1T,0.2T,\dots,1.0T$ here.}
\label{fig:packet}
\end{figure}

Let us then inspect the influence of the electric field (\ref{eq:electric-field}) to the flat bands. In the rotating frame Eq.~(\ref{eq:rotating-transformation}), the time average of Eq.~(\ref{eq:electric-field}) is given by
\begin{align}
\overline{U^\dagger(t)H_E(t)U(t)}&= \frac{E}{2\Omega}\sin\theta \sigma_z = \frac{E}{\Omega^2}p_y \sigma_z.
\end{align}
Again, in the same way as the 3D Weyl fermion in a static magnetic field, the flat band acquires a linear dispersion 
\begin{equation}
    \varepsilon_0=-\tau\frac{\Omega}{2}-\frac{E}{\Omega^2}p_y.
\end{equation}
This implies that the $\pi$-Landau level exhibits the homodyne Hall effect, \textit{i.e.}, it carries the DC Hall current under the AC electric field.

As we have obtained the explicit form of the wave function, here we discuss an intuitive picture for the $\pi$-Landau level states.
As we detail in Appendix~\ref{sec:appendix-LM}, the orbital angular momentum carried by the $\pi$-Landau level is given by
\begin{align}
\langle u_{\pm,n}(t)|\hat{L}_z|u_{\pm,n}(t)\rangle&=2n \cos(\Omega t),\\
\langle u_{0}(t)|\hat{L}_z|u_{0}(t)\rangle&= \cos(\Omega t),
\end{align}
where
\begin{equation}
    |u_{n,\pm}(t)\rangle=U(t)|v_{\pm,n}\rangle,\quad|u_{0}(t)\rangle=U(t)|v_{0}\rangle
\end{equation}
is the time-periodic part of the Floquet state in the original frame.
This expression coincides with that for the Landau level in a static magnetic field Eq.~(\ref{eq:LL-Lz}), when $\text{sgn}B$ is replaced into $\cos(\Omega t)$. This fact implies that the cyclotron motion of the electron becomes clockwise and counterclockwise in a periodic manner, synchronized with the AC field $B_z(t)=B\cos(\Omega t)$. 
This picture is consistent with the semiclassical picture shown in Sec.~\ref{sec:semiclassical}, where 
the trajectory of the electron takes the snake-like shape as shown in Fig.~\ref{fig:trajectory}.
In particular, as we show in Appendix~\ref{sec:appendix-LM}, the $y$ coordinate of the wave packet composed of the flat-band state $|u_0(t)\rangle$ also exhibits temporal oscillation $\langle y\rangle\propto\cos(\Omega t)$, which implies that the flat-band state constitutes the figure-of-eight shaped closed orbit (clockwise for $y<0$ and counterclockwise for $y>0$) as in the semiclassical case.

On the other hand, in contrast to the semiclassical case, the $\pi$-Landau level states utilizes a quantum nature in realizing the snake-like orbit. Namely, the wave function of the $\pi$-Landau level states can be interpreted as a superposition of the clockwise and counterclockwise cyclotron orbits, as
\begin{align}
 |u_{n,\pm}(t)\rangle  
 & =e^{i(\Omega/2)(x-\tau t)}\left(\cos\frac{\Omega t}{2}|v_{n,\pm}(+B)\rangle-i\tau\sin\frac{\Omega t}{2}|v_{n,\pm}(-B)\rangle\right),
\end{align}
where $|v_{n,\pm}(+B)\rangle=(\pm|n-1\rangle,|n\rangle)^{\text{T}}/\sqrt{2}$
and $|v_{n,\pm}(-B)\rangle=(|n\rangle,\pm|n-1\rangle)^{\text{T}}/\sqrt{2}$
are the static Landau level states for positive and negative magnetic
fields, respectively [See Eqs.~(\ref{eq:static-2D-state-positive}) and (\ref{eq:static-2D-state-negative})]. Here we have set $\theta=0$ for simplicity.

We can also calculate the expectation value of the orbital magnetic moment, which yields
\begin{align}
\langle u_{\pm,n}(t)|\hat{M}_z|u_{\pm,n}(t)\rangle& =\mp\frac{1}{2}\tau\sqrt{\frac{n}{B}}\frac{1}{\cos\theta}\cos(\Omega t),\\
\langle u_{0}(t)|\hat{M}_z|u_{0}(t)\rangle& =-\frac{\tan\theta}{2\Omega}\sin(\Omega t)\cos(\Omega t).
\end{align}
See Appendix~\ref{sec:appendix-LM} for details. As in the static Landau levels, the label $\pm$ for the Floquet states can be interpreted as the label specifying the electron and hole branches, and the magnetic moment has different signs for the two branches. 
On the other hand, as opposed to the static field case, the flat band state acquires the nonzero magnetic moment that oscillates with the frequency $2\Omega$.  The additional factor of $\sin(\Omega t)$ implies a resonant oscillation in the fraction of the electron and hole components.

Figure~\ref{fig:packet} compares the approximate solution Eq.~(\ref{eq:solution}) denoted as ``Ansatz" with the numerically obtained solution
for $B=2/l_0^{2}, p_y=0$.  
Here, for simplicity, we take a linear combination as
\begin{align}
|\tilde{\psi}_{0}(t)\rangle & =\frac{1}{\sqrt{2}}\left[U(t)|v_{0}\rangle|_{\theta=0}-iU(t)|v_{0}\rangle|_{\theta=\pi}\right]e^{i\tau\Omega t/2}\\
 & =\sqrt{2}\begin{pmatrix}\tau\sin\frac{\Omega x}{2}\sin\frac{\Omega t}{2}|0\rangle\\
\cos\frac{\Omega x}{2}\cos\frac{\Omega t}{2}|0\rangle
\end{pmatrix},
\end{align}
with which the wave function becomes real.
The good agreement implies that the intuitive picture discussed above is justified in the present choice of parameters.

We remark on the validity of the high-frequency expansion in the rotating frame Eq.~(\ref{eq:rotating-frame}).
The matrix element of oscillating terms in Eq.~(\ref{eq:rotating-frame}) acting on $|n\rangle$ is $\sim \sqrt{Bn}$, which implies that the high-frequency expansion is expected to give a good approximation
for the Landau levels with small $n$
when the field amplitude $B$ is small compared with $B_0=\Omega^2=4\pi^2 /l_0^{2}$, as numerically demonstrated above.
We discuss the breakdown of the high-frequency expansion in more detail in Sec.~\ref{sec:higher}.

\section{Symmetry and robustness}\label{sec:robustness}
While the time-averaged Hamiltonian in the rotating frame $\overline{H_{\text{rot}}(t)}$
captures the main feature around the Floquet Brillouin zone boundary $\varepsilon=\pm\Omega/2$,
its robustness is a highly-nontrivial issue.
In usual cases, the flat band state in the time-averaged Hamiltonian should be spoiled, 
when the perturbation (\textit{i.e.}, the higher-order terms in the high-frequency expansion) is taken into account. 
Why do the doubly-degenerate zeroth Landau levels remain flat and pinned to $\varepsilon=\Omega/2$
even in the presence of perturbation? 
In this section, we discuss the robustness of the flat band states in terms of the symmetry.

\subsection{Robustness of the flat band}
Let us first focus on the robustness of the flat band state.
The robust nature turns out to be inherent to the chiral symmetry in a dynamical sense, expressed as
\begin{equation}
\sigma_{z}H(p_y,-t)\sigma_{z}=-H(p_y,t),
\label{eq:chiral-symmetry}
\end{equation}
for $E=0$.
This relation represents the symmetry, in the sense that, 
when the Floquet state $|\psi_{\alpha}(p_y,t)\rangle=|u_{\alpha}(p_y,t)\rangle e^{-i\varepsilon_{\alpha}t}$
satisfies the time-dependent Schr\"odinger equation, $|\Gamma\psi_{\alpha}(p_y,t)\rangle=\sigma_{z}|\psi_{\alpha}(p_y,-t)\rangle$
is also the solution with quasienergy $-\varepsilon_{\alpha}$.

While the chiral symmetry makes the quasienergy
spectrum symmetric around $\varepsilon=0$, due to the periodicity in
quasienergy it must also be symmetric around $\varepsilon=\Omega/2\equiv-\Omega/2$ mod $\Omega$.
$\varepsilon=\Omega/2$ as well as $\varepsilon=0$ are the invariant
quasienergy under the chiral operation when the system is periodically
driven, which implies that the quasienergy must take either of these
values when the Floquet state is the simultaneous eigenstate of the
chiral operation $\Gamma$. Indeed, the approximate flat band state in the original
frame 
\begin{align}
U(t)|v_{0}\rangle e^{i\tau\Omega t/2}
 & =e^{i(\Omega/2)x\cos\theta}
  \begin{pmatrix}-i\tau\sin\frac{\Omega t}{2}e^{-i\theta/2}|0\rangle\\
  \cos\frac{\Omega t}{2}e^{i\theta/2}|0\rangle
  \end{pmatrix} 
\end{align}
is the chiral eigenstate with the eigenvalue of $\Gamma=-1$. An important
observation here is that 
the chiral eigenvalue $\Gamma$ takes the same value for the two choices of $\theta$ for the rotating frame Eq.~(\ref{eq:rotating-transformation}), which we have identified as an origin of the degeneracy in the previous section.
As any linear combination
of these states is also the chiral eigenstate, the quasienergy is expected to be pinned at $\varepsilon=\Omega/2$.

We can show that the chiral eigenstates are indeed stable against any perturbation $V$ respecting the chiral symmetry $\Gamma$, 
as long as matrix elements of $V$ between $\Gamma=+1$ and $\Gamma=-1$ states are negligible.
We provide an elementary derivation using the Sambe space formalism [Eq.~(\ref{eq:sambe})] in Appendix~\ref{sec:appendix-robustness}. 
Because there are no $\Gamma=+1$ state in the present energy scale of interest, the doubly-degenerate
flat band with $\Gamma=-1$ is quite stable. 

\begin{figure}[thb]
\centering
\includegraphics[width=\linewidth]{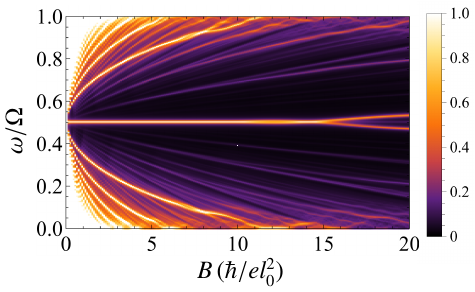}
\caption{Energy spectrum at $p_y=0$ as a function of magnetic field amplitude $B$.}
\label{fig:breakdown}
\end{figure}

We examine the robustness of the flat band numerically in Fig.~\ref{fig:breakdown}.
Here we plot the energy spectrum at $p_y=0$ against the field amplitude $B$. 
We can see that the flat band state survives up to $B\simeq 10/l_0^{2}$,
while it finally splits in the strong field regime. 
This observation implies that there are hidden $\Gamma=+1$ states in the higher energy scale, which we identify in Sec.~\ref{sec:higher}.

\subsection{Degeneracy of the Landau levels}

Let us move on to the double degeneracy of the flat band in terms of symmetry.
This degeneracy is actually related to the nonsymmorphic space-time symmetry, which is a dynamical symmetry peculiar to periodically-driven systems.
The nonsymmorphic space-time symmetry $G$ relevant to the double degeneracy is dictated for the present model as
\begin{equation}
\sigma_{x}H\left(-p_y,t+\frac{T}{2}\right)\sigma_{x}=H(p_y,t),
\end{equation}
with which $|G\psi_{\alpha}(p_y,t)\rangle=\sigma_{x}|\psi_{\alpha}(-p_y,t+T/2)\rangle$ 
is the Floquet state with quasienergy $\varepsilon_{\alpha}$ when $|\psi_{\alpha}(p_y,t)\rangle$ is the Floquet state (of $H(-p_y,t)$) with quasienergy $\varepsilon_{\alpha}$.

By combining the static antiunitary symmetry $K$ given by
\begin{equation}
H^\ast(p_y,t)=-H(p_y,t),
\end{equation}
we can indeed show in an elementary manner that the Floquet state with $\varepsilon=\Omega/2$ must be doubly-degenerate, in a similar way as the symmetry-enforced degeneracy of zero modes in class C;
Since the symmetry partner $|KG\psi_{\alpha}(p_y=0,t)\rangle=\sigma_{x}|\psi_{\alpha}^\ast(p_y=0,t+T/2)\rangle$ is the Floquet state with quasienergy $-\varepsilon_\alpha$, it leads to the degeneracy for $\varepsilon_\alpha=0,\Omega/2$ when this state is linearly-independent from the original state. By applying $KG$ once again, we obtain $|(KG)^2\psi_{\alpha}(p_y=0,t)\rangle=|\psi_{\alpha}(p_y=0,t+T)\rangle=|\psi_{\alpha}(p_y=0,t)\rangle e^{-i\varepsilon_\alpha T}$, 
which implies $(KG)^2=-1$ for $\varepsilon_\alpha=\Omega/2$. With this we can proof by contradiction that $|KG\psi_{\alpha}(p_y=0,t)\rangle$ for $\varepsilon_\alpha=\Omega/2$ is always linearly-independent from $|\psi_{\alpha}(p_y=0,t)\rangle$. While only the degeneracy at $p_y=0$ is shown here, the robustness due to the chiral symmetry ensures the degenerated state at $p_y\neq0$.

The degeneracy here is an analog of those at the Brillouin zone edge in nonsymmorphic crystals, and also has a close relation to the Floquet gap-dependent topological classification~\cite{Na2023} for the truncated Hamiltonian in the Sambe space. Indeed, if we consider the Sambe-space representation of the symmetry operation $KG$, we can check that $(KG)^2=+1$ ($-1$) for the truncation of the Sambe space with even (odd) numbers of sectors, which implies the gap-dependent degeneracy.

\subsection{Flat bands at higher harmonics}\label{sec:higher}
In this subsection, we reveal the presence of the chiral states whose spectral weight is peaked at $\omega=n\Omega/2$ with $n=3,5,\dots$, 
by generalizing the rotating frame Eq.~(\ref{eq:rotating-transformation}) to higher harmonics.
We introduce $U_n(t)$ here based on $U(t)$, as
\begin{align}
 U_{n} = e^{i(n\Omega/2)x\cos\theta_n}e^{-i\sigma_z\theta_n/2}e^{-i\tau(n\Omega/2)(1+\sigma_x)t},
\end{align}
with $\theta_{n}=\sin^{-1}[p_{y}/(n\Omega/2)]$. 
Then the Hamiltonian in the rotating frame $H_n(t)=U_{n}^{\dagger}(t)(H(t)-i\partial_{t})U_{n}(t)$ is obtained as 
\begin{align}
 H_{n}(t) &=-\tau\frac{n\Omega}{2}+\tau[\sigma_{x}\cos\theta_{n}-(\sigma_{x}^{+}e^{i\tau n\Omega t}+\text{H.c.})\sin\theta_{n}](-i \partial_{x})\nonumber\\
 & +\tau [\sigma_{x}\sin\theta_{n}+(\sigma_{x}^{+}e^{i\tau n\Omega t}+\text{H.c.})\cos\theta_{n}]Bx \cos\Omega t.\label{eq:rotating-frame-n}
\end{align}
Unlike the previous one with $n=1$, the time average of the Hamiltonian for $n>1$ is given just by the momentum operator, 
\begin{align}
\overline{H_{n}(t)} & =-\tau\frac{n\Omega}{2}+\tau\sigma_{x}\cos\theta_{n}(-i \partial_{x}),
\end{align}
and has no magnetic field term. 

As we show in Appendix~\ref{sec:appendix-higher}, the high-frequency expansion of Eq.~(\ref{eq:rotating-frame-n}) yields
a nonuniform static magnetic field term in higher orders.
Using the Brillouin-Wigner expansion~\cite{Mikami2016}, we obtain the effective Hamiltonian for $n=3$ as
\begin{align}
H_{\text{eff}} & =-\tau\frac{3\Omega}{2}+\tau\cos\theta_{3}\left[\sigma_{x}(-i \partial_{x})-\sigma_{y}B_3x^3\right]
\end{align}
with 
\begin{align}
B_3&=\frac{B^3}{32\Omega^2}(1-9\sin^{2}\theta_3)=\frac{B^3}{32\Omega^2}\left[1-\left(\frac{2p_y}{\Omega}\right)^2 \right].\label{eq:magnetic-field-3}
\end{align}
Let us focus on the case $|p_y|<\Omega/2$ with $B_3>0$ for simplicity. 
We can obtain the zero-energy eigenstate as a kernel of the matrix element, as
\begin{align}
|v_{3;0}\rangle\propto\int dxe^{-B_{3}x^{4}/4}\begin{pmatrix}|x\rangle\\
0
\end{pmatrix}.
\end{align}
This is nothing but a chiral eigenstate with $\Gamma=+1$, which lead to the level splitting when hybridized with the $\pi$-Landau level $|v_0\rangle$.

In summary, the flat band state obtained as an eigenstate of  $\overline{H_{\text{rot}}(t)}$ is shown to be robust due to the chiral symmetry. The robustness is assured when the perturbation mixing the chiral eigenstates with $\Gamma=+1$ and $\Gamma=-1$ is vanishingly small. We identify the hidden $\Gamma=+1$ states in the higher energy scale, which are decoupled from the flat band states with $\Gamma=-1$ in the high-frequency (weak-field) limit, but are hybridized in the strong field regime.

\begin{figure*}[tbh]
\centering
\includegraphics[width=\linewidth]{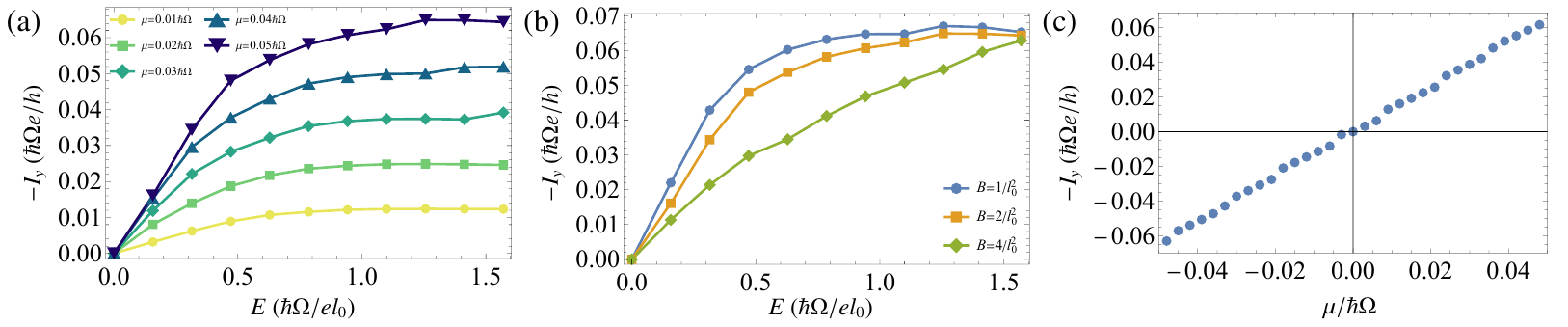}
\caption{(a,b) DC Hall current as a function of AC electric field amplitude $E_0$ for  $B=2\hbar/el_0^2$ (a) and $\mu=0.05\hbar\Omega$ (b). (c) DC Hall current as a function of chemical potential for $E=0.94\hbar\Omega/el_0$.}
\label{fig:hallcurrent}
\end{figure*}

\section{Homodyne Hall current}\label{sec:current}
In this section, we study dynamical transport properties of the 2D Dirac electrons in two AC-fields. 
Generically, when two fields with frequencies $\Omega_1$ and $\Omega_2$ are applied to an electronic system, we expect generation of a heterodyne current with a frequency given by their difference  $|\Omega_1-\Omega_2|$. 
When the frequencies are equal, the output can be a DC current referred to as a homodyne current.  
While the standard mechanisms known for the generation of heterodyne and homodyne currents are perturbative, here we find a DC homodyne current as a nonperturbative effect. 
To be specific, we consider the current generated by the two fields
\begin{equation}
A_y=B\cos(\Omega t)x \quad \mbox{and}\quad 
A_x=\frac{eE}{\Omega}\sin(\Omega t),
\end{equation}
where their effects on the spectral properties have been discussed in the previous sections. 
The combination of the two fields breaks time-reversal symmetry, and we guess that there is a current generated by the fields. 
In particular, we expect a DC-current flowing perpendicular to the $E$-field
which we refer to as the homodyne Hall current $I_y$ in 2D Dirac electron. 

Here, we evaluate the DC-current flowing in the system assuming that the system is weakly coupled to a local fermion bath. 
In this case, the DC-current of the $\alpha$-th Floquet eigenstate is given as the group velocity $j_{y}^\alpha=-\partial_{p_y}\varepsilon_\alpha$,
and their occupation $f_\alpha$ is described in Eq.~(\ref{eq:occupation}). 
Thus the total DC-current is given by $I_y=\int \frac{dp_y}{2\pi}\sum_\alpha j_y^\alpha f_\alpha $ (See details for Appendix~\ref{sec:appendix-current}). 

We show the numerical result of the homodyne Hall current for the 2D Dirac fermion in Fig.~\ref{fig:hallcurrent}
for several values of chemical potential $\mu$ and magnetic field parameter $B$. 
Here, we have assumed that the system is charge-neutral and has vanishing current at $\mu=0$. We integrate the current contribution in $p_y\in[-\Omega/2,\Omega/2]$.
As the field amplitude is increased, the Hall current is initially proportional to the AC-electric field strength $E$ 
and then saturates to a value that is dependent on the chemical potential $\mu$
but not on the AC-magnetic field $B$ [See Figs.~\ref{fig:hallcurrent}(a,b)]. 
In fact, as plotted in Fig.~\ref{fig:hallcurrent}(c), we observe that the saturated value of the homodyne Hall current is well represented by 
\begin{equation}
I_y=-\frac{e}{h}\mu\,C
\label{eq:dchall}
\end{equation}
with $C\sim1$, per valley and spin (in the current calculation the spin degeneracy is unity). 
This result is very different from nonrelativistic electrons with parabolic bands~\cite{Oka2016}
where the homodyne Hall effect is proportional to the strength of the applied AC-field $E$. 
How can we understand this exotic behavior? Why is the current generated by the two fields no longer depend on $B$ nor $E$? 
This anomalous behavior of the homodyne Hall current can be related to the current in 3D Weyl fermions in magnetic fields [Eq.~(\ref{eq:CMEcurrent})]. 
Indeed, the Floquet effective Hamiltonian that governs the coarse-grained dynamics given in Eq.~(\ref{eq:HFchiral})
is analogous to the 3D Weyl fermion in magnetic fields projected to zero transverse momentum $p_z=0$
and the tilted Floquet spectrum [see Figs.~\ref{fig:linear}(d) and \ref{fig:spectrum}(c)] corresponds to the chiral Landau level. 
There are two chiral bands at zero energy for the two valleys and are both nondegenerate. 
Since the chiral current (per valley and spin) in the 3D Weyl fermion is given by 
$I=L^2B\mu e^2/h^2$ and carried by the chiral band with degeneracy $D=L^2Be/h$, 
we can deduce the anomalous current for nondegenerate chiral band as $I/D=\mu e/h$, which agrees with Eq.~(\ref{eq:dchall}).

\begin{figure*}
\centering
\includegraphics[width=\linewidth]{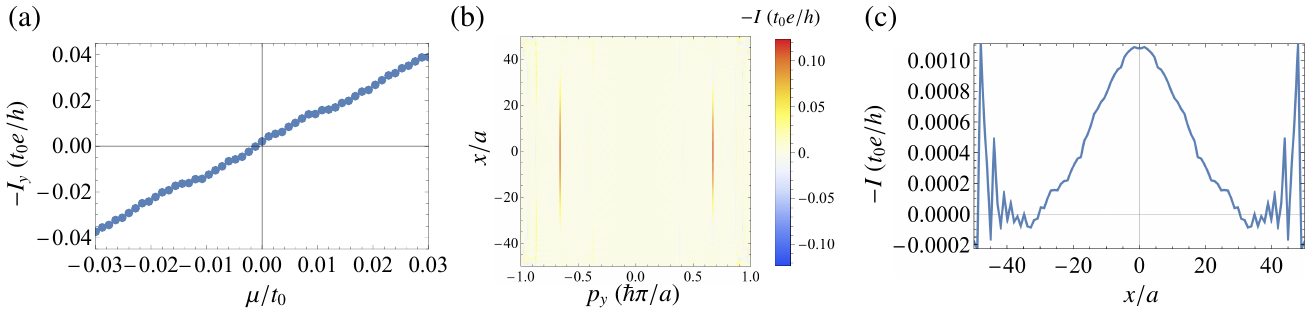}
\caption{(a) Hall current in honeycomb lattice model for $E=0.05\hbar\Omega/ea$, as a function of chemical potential $\mu$. 
(b) Current resolved in spatial position $x$ and momentum $p_y$, for $\mu=0.048\hbar\Omega$. (c) Spatial profile of the Hall current for $\mu=0.048\hbar\Omega$ [\textit{i.e.}, momentum integration of (b)].
}
\label{fig:hallcurrent2}
\end{figure*}

An interesting point here is that the chiral bands for two valleys have the same chirality and are both right movers, with which we expect that the cancellation of the current known in the 3D Weyl fermion can be circumvented in the present case. In order to confirm this, we numerically computed the DC Hall current for the lattice model Eq.~(\ref{eq:lattice-model}) with two valleys. We plot the chemical potential dependence of the Hall current for $E=0.05\Omega/a$ in Fig.~\ref{fig:hallcurrent2}(a). Although the slope of the current $C$ is slightly degraded, the DC Hall current survives even if we add up the contribution of two valleys as well as that of other bands and edges.
We further investigate the origin of the Hall current in Figs.~\ref{fig:hallcurrent2}(b) and (c), where we compute the Hall current resolved in position $x$ and momentum $p_y$. For clarity, here we show the difference between current at $\mu=0.048\Omega$ and that at $\mu=0$. As shown in Figs.~\ref{fig:hallcurrent2}(b), the current profile has sharp peaks at the momentum corresponding to K and K$'$, where the chiral bands are located.  These peaks are spatially localized and centered at $x=0$, and has dominant contribution, as seen in the momentum-integrated value displayed in  Figs.~\ref{fig:hallcurrent2}(c). 

Finally, let us explain how to convert the parameters used in the calculation to experimental values. 
In this paper, the field strength are scaled using the parameter $l_{0}=2\pi v_{F}/\Omega$ that has the dimension of length. 
For example, the AC-magnetic field $B$ and AC-electric field $E$
uses the unit of $(\hbar/el_{0}^{2})$ and $(\hbar\Omega/el_{0})$ respectively. 
The calculation for the lattice model in Fig.~\ref{fig:spectrum} employs $\hbar\Omega=0.6t_0$,
which results in preparing $B$ in units of $\hbar/el_{0}^{2}\sim \SI{100}{\tesla}$
and $E$ in units of $\hbar\Omega/el_{0}\sim \SI{8}{\mega\volt/\cm}$ obtained by $l_0\sim\SI{2}{\nm}$ with $t_0\sim\SI{3}{\eV}$ and $v_F\sim c/300$ for graphene. 
If we are to scale down the frequency of the AC-fields, then the field strengths $B$ and $E$ scale down quadratically $\propto \Omega^2$. 
For example, if we go down to the THz regime $\hbar \Omega\sim\SI{4}{\meV}$ which is 500 times smaller than the above 
parameter sets, the fields become $500^2$ times smaller, \textit{i.e.} $B\sim\SI{6E-4}{\tesla}$
and $E\sim \SI{40}{\volt/\cm}$ with $l_0\sim\SI{1}{\um}$.

If we are interested in observing the chiral anomaly-induced homodyne current, the sample should be larger than the 
magnetic length of the chiral band. 
The effective magnetic field for the chiral band is given by $B_{\text{eff}}=(ev_{F}/\hbar)BE/\Omega^{2}\sim(2\pi)^{-1}\hbar/el_{0}^{2}$ 
resulting in the magnetic length $l_B=\sqrt{\hbar/eB_\text{eff}}\sim \sqrt{2\pi}l_0$ using the above unit field strengths $B,E$.
For the THz case, the magnetic length becomes $l_B\sim \SI{3}{\um}$. 
The magnitude of the homodyne current with $\mu\sim0.05\hbar\Omega$ computed in Fig.~\ref{fig:hallcurrent} in the THz case is translated into $I_y\sim\SI{8}{\nA}$, which may be experimentally detectable.

\section{Conclusion}\label{sec:discussion}
In this paper, we have investigated the spectral and transport properties of the 2D Dirac electrons in the presence of the AC-magnetic field.
We find that the doubly-degenerate flat bands ($\pi$-Landau levels) appear at $\omega=\pm\Omega/2$ under the AC-magnetic field, and the nondegenerate chiral band appears when we additionally apply an AC-electric field with the same frequency.

To describe these characteristic eigenstates, we construct the effective Floquet Hamiltonian. After an appropriate unitary transformation, the $\pi$-Landau level turns out to be described as the chiral Landau level. The obtained wave function can be viewed as a resonant state between the clockwise and counterclockwise cyclotron orbits, where the transition between two branches is synchronized with the driving period. Due to the chiral nature of the $\pi$-Landau level, the flat dispersion at $\omega=\Omega/2$ is shown to be robust against perturbation, although the flatness is broken down when the applied magnetic field becomes strong and the $\pi$-Landau levels hybridize with the high-energy states.

The chiral band under the additional AC electric field is obtained by the high-frequency expansion, where the resultant effective Hamiltonian can be viewed as the 3D Weyl fermion in a static magnetic field, projected onto the $p_z=0$ plane. An interesting point here is that the static magnetic field in the effective Hamiltonian is chirality-dependent, and the group velocity of the chiral bands have the same sign for two valleys. This implies that, unlike the 3D Weyl fermions, the anomalous current associated with the chiral bands does not cancel out. We demonstrate that the homodyne Hall current under the AC magnetic and electric fields ($B$ and $E$) is proportional to the chemical potential, and saturates to a value independent of $B$ and $E$.
The prefactor to the Hall current is consistent with the anomalous contribution estimated with the equilibrium distribution function.

Recently, the role of quantum geometry in flat band systems is intensively studied in fractional quantum Hall effect~\cite{Wang2021,Ledwith2023} and superconductivity~\cite{Peotta2015}.
The geometric characterization of the present system hosting a new type of flat bands is 
an interesting future problem. 
To do so, we need to extend the concept of quantum geometry 
to dynamical processes, and there has already been an interesting work~\cite{Martin2017} in this direction. 
We hope our findings will provide another solid example 
that will guide us to understand the dynamical extension of quantum geometry.

\acknowledgments
The authors appreciate the fruitful discussions with K. Saha, T. Nag, 
A. Mitra, K. Richter, A. Cavalleri, and H. Hirori. 
This work is supported by 
JSPS KAKENHI (No. 23H04865, No.23K22487, No.23K25837, No. 23K25816, No. 25K07219), MEXT, Japan, 
and JST CREST Grant No. JPMJCR19T3, Japan.

\appendix

\section{Orbital angular momentum and orbital magnetic moment of the $\pi$-Landau level}\label{sec:appendix-LM}

In this appendix, we provide a derivation of the expectation value
of the orbital angular momentum and the orbital magnetic moment for
the $\pi$-Landau levels. In Sec.~\ref{sec:preliminaries-2d}, we have shown the same
caluclation for the usual Landau level under the static magnetic field.
There, we have circumvented the calculation involving the position
operator $y$ using $\langle v|[O,H]|v\rangle=\langle v|[O,H^{2}]|v\rangle=0$
with $|v\rangle$ being the energy eigenstates. On the other hand,
since the $\pi$-Landau levels are the Floquet states but not the
eigenstate of the time-dependent Hamiltonian $H(t)$, we cannot use
the same trick and need to evaluate the matrix element of $y$ in
a direct manner. Since the position expectation value of $y$ is not
well-defined for the $\pi$-Landau levels extended along the $y$ direction,
here we consider a wave packet consisting of the $\pi$-Landau level
and take the large-width limit in the final step.

Since the wave function of the $\pi$-Landau level for a fixed
momentum $p_{y}$ is given by $|\psi_{n,\pm}(t)\rangle=U(t)|v_{n,\pm}(t)\rangle e^{-i\varepsilon_{n,\pm}t}$
with Eqs.~(\ref{eq:rotating-transformation}), (\ref{eq:solution}) and (\ref{eq:piLL-dispersion}), we can define the wave
packet of the $\pi$-Landau level $|w_{n,\pm}\rangle$ by
\begin{equation}
|w_{n,\pm}\rangle=\int\frac{dp_{y}}{2\pi}C(p_{y})|\psi_{n,\pm}(t)\rangle\otimes|p_{y}\rangle,
\end{equation}
where $|p_{y}\rangle=\int dye^{ip_{y}y}|y\rangle$ is the momentum
eigenstate along the $y$ direction. Here, the momentum distribution of
the wave packet $C(p_{y})$ has nonzero values only around some momentum
$p_{y}=p_{c}$. Then the action of the position operator $y$ on the
wave packet can be calculated using $y|p_{y}\rangle=-i\partial_{p_{y}}|p_{y}\rangle$
and integration by part, as 
\begin{align}
y|w_{n,\pm}\rangle & =\int\frac{dp_{y}}{2\pi}i\partial_{p_{y}}C(p_{y})|\psi_{n,\pm}(t)\rangle\otimes|p_{y}\rangle\nonumber \\
 & +\int\frac{dp_{y}}{2\pi}C(p_{y})\frac{2}{\Omega\cos\theta}i\partial_{\theta}|\psi_{n,\pm}(t)\rangle\otimes|p_{y}\rangle,
\end{align}
where we have used the fact that $|\psi_{n,\pm}(t)\rangle$ depends
on $p_{y}$ through $\theta=\sin^{-1}(2p_{y}/\Omega)$. The $\theta$
derivative is given by
\begin{align}
i\partial_{\theta}|\psi_{n,\pm}(t)\rangle & =\left(\frac{\Omega x}{2}\sin\theta+\frac{1}{2}\sigma_{z}\mp\tau\sqrt{Bn}t\sin\theta\right)|\psi_{n,\pm}(t)\rangle.
\end{align}

With these, we obtain the center-of-mass position of the wave packet $y_{c}=\langle w_{n,\pm}|y|w_{n,\pm}\rangle$ 
as 
\begin{align}
y_{c} 
 & =-\int\frac{dp_{y}}{2\pi}|C|^{2}\left(\partial_{p_{y}}\arg C\pm\tau\frac{2\sqrt{Bn}}{\Omega}t\tan\theta+\delta_{n,0}\frac{\cos\Omega t}{\Omega\cos\theta}\right)
\end{align}
using $U^{\dagger}(t)\sigma_{z}U(t)=\sigma_{z}\cos\Omega t+\tau\sigma_{y}\sin\Omega t$.
Here we have combined the result for $|v_{0}\rangle$ as the $n=0$
case. In the same manner, we can calculate the expectation value of
$yp_{x}$ and $y\sigma_{x}$ using $U^{\dagger}(t)p_{x}U(t)=p_{x}+(\Omega/2)\cos\theta$ and 
$U^{\dagger}(t)\sigma_{x}U(t)=\sigma_{x}\cos\theta-(\sigma_{y}\cos\Omega t-\tau\sigma_{z}\sin\Omega t)\sin\theta$,
as
\begin{align}
\langle w_{n,\pm}|yp_{x}|w_{n,\pm}\rangle & =-\int\frac{dp_{y}}{2\pi}|C|^{2}\partial_{p_{y}}(\arg C)\frac{\Omega}{2}\cos\theta\nonumber \\
 & -\int\frac{dp_{y}}{2\pi}|C|^{2}\left(\pm\tau\sqrt{Bn}t\sin\theta+\frac{1}{2}\delta_{n,0}\cos\Omega t\right),\\
\langle w_{n,\pm}|y\sigma_{x}|w_{n,\pm}\rangle & =\int\frac{dp_{y}}{2\pi}|C|^{2}\tau\delta_{n,0}\partial_{p_{y}}(\arg C)\sin\theta\sin\Omega t\nonumber \\
 & \pm\int\frac{dp_{y}}{2\pi}|C|^{2}\sqrt{\frac{n}{B}}\left(\cos\theta-\frac{1}{\cos\theta}\right)\cos\Omega t.
\end{align}

Let us consider the limit where the width of the peak in $|C(p_{y})|^{2}$
is infinitesimally small, so that the integral can be approximated
as
\begin{equation}
\int\frac{dp_{y}}{2\pi}|C(p_{y})|^{2}f(p_{y})\to f(p_{c})\int\frac{dp_{y}}{2\pi}|C(p_{y})|^{2}=f(p_{c}).
\end{equation}
This corresponds to the limit where the width of the wave packet in
the real space is infinitely large, and the wave packet converges to a plane wave. In this limit, the quantities calculated
above can be decomposed into the center-of-mass part and the remainder
as
\begin{align}
\langle w_{n,\pm}|yp_{x}|w_{n,\pm}\rangle & \to y_{c}\langle w_{n,\pm}|p_{x}|w_{n,\pm}\rangle,\\
\langle w_{n,\pm}|y\sigma_{x}|w_{n,\pm}\rangle & \to y_{c}\langle w_{n,\pm}|\sigma_{x}|w_{n,\pm}\rangle-\delta_{n,0}\frac{\tau}{2\Omega}\tan\theta\sin2\Omega t\nonumber \\
 & \pm\sqrt{\frac{n}{B}}\left(\cos\theta-\frac{1}{\cos\theta}\right)\cos\Omega t.
\end{align}
The expectation values of $x(p_{y}+Bx\cos\Omega t)$ and $x\sigma_{y}$
can be calculated in a similar manner as Eqs.~(\ref{eq:LL-Lz}) and (\ref{eq:LL-Mz}),
using $U^{\dagger}(t)\sigma_{y}U(t)=\sigma_{x}\sin\theta+(\sigma_{y}\cos\Omega t-\tau\sigma_{z}\sin\Omega t)\cos\theta$.
Then, $\langle\hat{L}_{z}\rangle$ and $\langle\hat{M}_{z}\rangle$
excluding the center-of-mass contribution can be obtained as
\begin{align}
\langle\hat{L}_{z}\rangle & =\langle w_{n,\pm}|[x(p_{y}+Bx\cos\Omega t)-(y-y_{c})p_{x}]|w_{n,\pm}\rangle\\
 & =(2n+\delta_{n,0})\cos\Omega t,\\
\langle\hat{M}_{z}\rangle & =-\frac{1}{2}\tau\langle w_{n,\pm}|[x\sigma_{y}-(y-y_{c})\sigma_{x}]|w_{n,\pm}\rangle\\
 & =\mp\frac{1}{2}\tau\sqrt{\frac{n}{B}}\frac{1}{\cos\theta}\cos\Omega t-\delta_{n,0}\frac{\tan\theta}{4\Omega}\sin2\Omega t.
\end{align}

\section{Robustness of chiral eigenstates}\label{sec:appendix-robustness}
In this appendix, we show that the eigenstates of the chiral operator
$\Gamma$ is robust against perturbation that mixes them with other
eigenstates. First, let us consider a static system with chiral symmetry,
$\{H,\Gamma\}=0$, $\Gamma^{2}=1$. We define the energy eigenstates
as $|\chi_{i}^{\sigma}\rangle$ and $|\xi_{i}^{\sigma}\rangle$ with
$\sigma=\pm1$, where $|\chi_{i}^{\sigma}\rangle$ denotes chiral
eigenstates satisfying 
\begin{equation}
H|\chi_{i}^{\sigma}\rangle=0,\quad\Gamma|\chi_{i}^{\sigma}\rangle=\sigma|\chi_{i}^{\sigma}\rangle,
\end{equation}
 while $|\xi_{i}^{\sigma}\rangle$ denotes usual eigenstates satisfying
\begin{equation}
H|\xi_{i}^{\sigma}\rangle=\sigma E_{i}|\xi_{i}^{\sigma}\rangle,\quad\Gamma|\xi_{i}^{\sigma}\rangle=|\xi_{i}^{-\sigma}\rangle.
\end{equation}

Let us consider a perturbation respecting the chiral symmetry, denoted as $V$. The symmetry relation $\{V,\Gamma\}=0$
leads to constraints on the matrix elements, 
\begin{gather}
\langle\chi_{i}^{\sigma}|V|\chi_{j}^{\sigma}\rangle=0,\\
\langle\xi_{i}^{-\sigma}|V|\chi_{j}^{\sigma}\rangle=-\sigma\langle\xi_{i}^{\sigma}|V|\chi_{j}^{\sigma}\rangle,\\
\langle\xi_{i}^{\sigma}|V|\xi_{j}^{\sigma^{\prime}}\rangle=-\langle\xi_{i}^{-\sigma}|V|\xi_{j}^{-\sigma^{\prime}}\rangle.
\end{gather}
Under these constraint, we can decompose the perturbation $V$ into
\begin{equation}
V=V_{\chi}+V_{\xi}+\sum_{j\sigma}V_{j\sigma},
\end{equation}
where $V_{\chi}$ has nonzero matrix elements only for $\langle\chi_{i}^{\sigma}|V_{\chi}|\chi_{j}^{-\sigma}\rangle$,
and $V_{\xi}$ does only for $\langle\xi_{i}^{\sigma}|V|\xi_{j}^{\sigma^{\prime}}\rangle$.
The last term mixing $\chi$ and $\xi$ is composed of 
\begin{align}
V_{j\sigma} & =\sum_{i}\alpha_{ij\sigma}(|\xi_{j}^{\sigma}\rangle-\sigma|\xi_{j}^{-\sigma}\rangle)\langle\chi_{i}^{\sigma}|+\text{H.c.}\label{eq:chiral-perturbation}
\end{align}

Evidently, $V_{\chi}$ leads to a gap opening for the chiral eigenstates
(thus washes out the robustness), while $V_{\xi}$ does not affect
the chiral eigenstates. A nontrivial consequence of the chiral symmetry
is the robustness of the chiral eigenstate against mixing perturbation
$V_{j\sigma}$. Indeed, for a fixed $j$ and $\sigma$, we can explicitly
constract a (non-orthogonal) set of new chiral eigenstates as
\begin{align}
|\chi_{i}^{\sigma}\rangle^{\prime} & =|\chi_{i}^{\sigma}\rangle-\frac{\alpha_{ij\sigma}}{E_{j}}(\sigma|\xi_{j}^{\sigma}\rangle+|\xi_{j}^{-\sigma}\rangle),
\end{align}
which satisfies $\Gamma|\chi_{i}^{\sigma}\rangle^{\prime}=\sigma|\chi_{i}^{\sigma}\rangle^{\prime}$.
We can check that this state is zero-energy eigenstate of $H+V_{j\sigma}$, using the
relation
\begin{equation}
V_{j\sigma}(\sigma|\xi_{j}^{\sigma}\rangle+|\xi_{j}^{-\sigma}\rangle)=0,
\end{equation}
 as
\begin{align}
(H+V_{j\sigma})|\chi_{i}^{\sigma}\rangle^{\prime} & =V_{j\sigma}|\chi_{i}^{\sigma}\rangle-\alpha_{ij\sigma}(|\xi_{j}^{\sigma}\rangle-\sigma |\xi_{j}^{-\sigma}\rangle)=0.
\end{align}
Repeating this construction (with orthonormalization) for all $j,\sigma$ yields the set of the chiral
eigenstates for $H+\sum_{j\sigma}V_{j\sigma}$. Therefore, when $V_{\chi}$
vanishes (or negligibly small) for some reason, the existence of the chiral eigenstates are unchanged even in the presence of a perturbation.

Let us then extend the argument above to periodically-driven systems, by formulating the symmetry relation in the Sambe space representation [see Eq.~(\ref{eq:sambe})]. When
the system has the chiral symmetry dictated as Eq.~(\ref{eq:chiral-symmetry}), we
can construct the chiral operator in the Sambe space as
\begin{equation}
\Gamma=\begin{pmatrix} &  &  &  & \iddots\\
 & O &  & \sigma_{z}\\
 &  & \sigma_{z}\\
 & \sigma_{z} &  & O\\
\iddots
\end{pmatrix},
\end{equation}
where each block stands for the Fourier components of real-time operators (tensor product with the identity operator in the $x$ space is omitted here). 
The nonlocal transformation from $t$ to $-t$ is easily handled and expressed with the anti-diagonal structure here, which is an advantage of the Sambe space formalism.
This operator obeys the chiral symmetry relation $\{\mathcal{H},\Gamma\}=0$, $\Gamma^2=1$, as in usual static systems, where $\mathcal{H}$ is the matrix in the left-hand side of Eq.~(\ref{eq:sambe}).

We can apply the argument above to Eq.~(\ref{eq:sambe}), and show that zero-energy
eigenstates in the Sambe space is robust against the perturbation
of the form Eq.~(\ref{eq:chiral-perturbation}). Here, note that the eigenstates with
eigenenergy $\pm\Omega/2,\pm\Omega,\pm3\Omega/2,\dots$ are not described
as the eigenstates of $\Gamma$ in the Sambe space, while they should be invariant under the chiral operation in
the real-time representation. 
We can capture the chiral nature of such eigenstates by additional
symmetry relations dictated as
\begin{align}
\left\{\mathcal{H}-\frac{N\Omega}{2},(M^{-})^{N}\Gamma\right\}  & =0,\quad\left[(M^{-})^{N}\Gamma\right]^{2}=1,
\end{align}
where $M^{-}$ is the Sambe space representation of $e^{i\Omega t}$,
expressed as
\begin{equation}
M^{-}=\begin{pmatrix}
\ddots & \ddots \\
 & O & I & & \\
 & & O & I & \\
 & & & O & \ddots\\
 & & & & \ddots
\end{pmatrix}
\end{equation}
with $I$ being the identity operator in $x$ and $\sigma$.
Since a general time-periodic perturbation satisfies $[V,M^-]=0$, 
$\{V,(M^{-})^{N}\Gamma\}=0$ always holds when $\{V,\Gamma\}=0$.
Namely, we can apply the argument above to the additional symmetries $(M^{-})^{N}\Gamma$ as well, and the eigenstates of $\mathcal{H}$ with $\varepsilon=N\Omega/2$ (\textit{i.e.}, the zero-energy eigenstates of $\mathcal{H}-N\Omega/2$) are also shown to be
robust against perturbation. In particular, $N=1$ 
describes nontrivial chiral eigenstates with $\varepsilon\not\equiv0$ mod $\Omega$, 
which corresponds to the $\pi$-Landau level in the present study.

\section{Derivation of the effective Hamiltonian for higher harmonic flat bands}\label{sec:appendix-higher}
In this appendix, we perform the high-frequency expansion of Eq.~(\ref{eq:rotating-frame-n}). 
For the present case with general $n$, 
the Brillouin-Wigner theory is suitable for deriving the static effective Hamiltonian. 
In the multi-root formalism of the Brillouin-Wigner expansion~\cite{Mikami2016},
the $N$-th order term with respect to $1/\Omega$ is given by [see Eq.~(14) of Ref.~\onlinecite{Mikami2016}]
\begin{equation}
H_{\text{BW}}^{(N)}=\sum_{n_1,n_2,\dots,n_N\neq0}\frac{H_{0,n_{1}}(\prod_{j=1}^{N-1}H_{n_{j},n_{j+1}})H_{n_{N},0}}{\prod_{j=1}^{N}(\varepsilon+n_{j}\Omega)},\label{eq:BW}
\end{equation}
where $H_{m,n}=H_{m-n}$ is the Fourier component of the Hamiltonian written as a matrix element in the Sambe space, and $\varepsilon$ is the eigenvalue of $H_{\text{BW}}$ that should be determined self-consistently.
Note that the constant energy shift of $O(\Omega)$ in Eq.~(\ref{eq:rotating-frame-n}) should be dropped in order to obtain $1/\Omega$ series, and then $\varepsilon=O(\Omega^0)$.

Among the terms of Eq.~(\ref{eq:BW}), that with the numerator composed of even numbers of $\sigma_x^\pm$ results in the $\sigma_x$ component, which can be absorbed by an $x$-dependent phase factor [as in the momentum shift performed with $V_1$ in Eq.~(\ref{eq:VHV})] and thus irrelevant here.
The static magnetic field term as a $\sigma_y$ component should consist of odd numbers of $\sigma_x^\pm$ perturbations.  
Since $(\sigma^\pm_x)^2=0$, such the term turns out to appear first at $N=n-1$,
and the numerator of Eq.~(\ref{eq:BW}) must be composed of $H_{\pm\tau n-\tau}\propto\sigma_{x}^{\mp}$
and $H_{-\tau}\propto\sigma_{x}$ for $(H_{\text{BW}}^{(n-1)})_{\downarrow,\uparrow}$. 
In this situation, the value of $n_j$ is restricted to 
$\tau j$ or $\tau(j-n)$, depending on the spin configuration.
By rewriting the $n_j$ summation in a $2\times2$ matrix product form, 
we obtain an explicit expression for the spin-flipping term as
\begin{widetext}
\begin{align}
H_{\text{BW}}^{(n-1)} & =\left(\frac{\tau Bx}{2}\right)^{n}\left\{ \begin{pmatrix}\sin\theta_{n} & \cos\theta_{n}\\
\cos\theta_{n} & -\sin\theta_{n}
\end{pmatrix}\prod_{j=1}^{n-1}\left[\begin{pmatrix}\varepsilon+\tau(j-n)\Omega & 0\\
0 & \varepsilon+\tau j\Omega
\end{pmatrix}^{-1}\begin{pmatrix}\sin\theta_{n} & \cos\theta_{n}\\
\cos\theta_{n} & -\sin\theta_{n}
\end{pmatrix}\right]\right\} _{\downarrow,\uparrow}\sigma_x^-+\text{H.c.}
\end{align}
\end{widetext}
In the leading-order evaluation with respect to $1/\Omega$, we can drop $\varepsilon$ in the energy denominator.
Then we obtain Eq.~(\ref{eq:magnetic-field-3}) for $n=3$.
For general $n$, we can simplify the matrix product above when $\theta_n=0$ ($p_y=0$). 
For this case, we can show that the above term vanishes for even $n$, while for odd $n$ with $n=2k+1$ we obtain 
\begin{equation}
H_{\text{BW}}\simeq-\tau\frac{n\Omega}{2}+\tau \sigma_{x}(-i\partial_{x})+(-1)^k\tau\sigma_{y}\frac{2\Omega}{(k!)^{2}}\left(\frac{Bx}{4\Omega}\right)^{n}.
\end{equation}
as an effective Hamiltonian of Eq.~(\ref{eq:rotating-frame-n}).
The zero-energy state can be obtained as a kernel of $(H_{\text{BW}})_{\uparrow,\downarrow}$ for even $k$ [$(H_{\text{BW}})_{\downarrow,\uparrow}$ for odd $k$],
which implies that it is a chiral eigenstate with $\Gamma=(-1)^{k+1}$.

\section{Derivation of DC current in the steady state}\label{sec:appendix-current}
Here we derive the expression for the DC current flowing in the steady
state described by Eq.~(\ref{eq:lesser}). The current expectation value
$I_{y}$ in terms of the nonequilibrium Green function is computed
as
\begin{align}
I_{y}(t) & =-e\int\frac{dp_{y}}{2\pi}\text{Tr}\left[-iG^{<}(t,t)\frac{\partial H(t)}{\partial p_{y}}\right]\\
 & =-\frac{e}{\hbar}\int\frac{dp_{y}}{2\pi}\sum_{\alpha\beta}f_{\alpha\beta}\langle u_{\beta}(t)|\frac{\partial H(t)}{\partial p_{y}}|u_{\alpha}(t)\rangle e^{-i(\varepsilon_{\alpha}-\varepsilon_{\beta})t/\hbar}.
\end{align}
Since the expression is expanded in terms of Floquet states satisfying
$i\hbar\partial_{t}[|u_{\alpha}(t)\rangle e^{-i\varepsilon_{\alpha}t/\hbar}]=H(t)|u_{\alpha}(t)\rangle e^{-i\varepsilon_{\alpha}t/\hbar}$,
we can rewrite the above expression as
\begin{align}
I_{y}(t) & =-\frac{e}{\hbar}\int\frac{dp_{y}}{2\pi}\sum_{\alpha} f_{\alpha\alpha}\frac{\partial\varepsilon_{\alpha}}{\partial p_{y}}\nonumber\\
&-ie\int\frac{dp_{y}}{2\pi}\sum_{\alpha\beta}f_{\alpha\beta}\frac{\partial}{\partial t}\left[\langle u_{\beta}(t)|\frac{\partial}{\partial p_{y}}|u_{\alpha}(t)\rangle e^{-i(\varepsilon_{\alpha}-\varepsilon_{\beta})t/\hbar}\right] .
\end{align}
Since $f_{\alpha\beta}$ given in Eq.~(\ref{eq:occupation}) is time-independent
in the $\Gamma\to0^{+}$ limit, the second term vanishes in time average.
Thus the total DC current is given by 
\begin{align}
\overline{I_{y}(t)} & =-\frac{e}{\hbar}\int\frac{dp_{y}}{2\pi}\sum_{\alpha}f_{\alpha\alpha}\frac{\partial\varepsilon_{\alpha}}{\partial p_{y}}.
\end{align}

\section{Numerical details}\label{sec:appendix-numerics}

\begin{figure}[t]
\centering
\includegraphics[width=\linewidth]{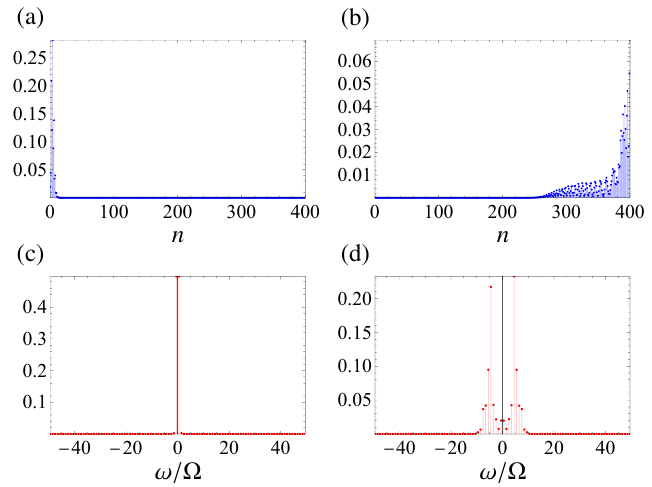}
\caption{Validation of the truncation scheme for the harmonic-oscillator basis and the Floquet-frequency cutoff used in the numerical calculations.
(a), (b) Distribution of the harmonic-oscillator modes, $\sum_{\sigma=\uparrow,\downarrow}\sum_m |\langle n,\sigma|u_{\alpha,m}\rangle|^2$, for representative Floquet states of the linearized Hamiltonian with $B=\hbar/el_0^2$, $E=0$, and $p_y=0.2\hbar\Omega/v_F$. The weight with $n\ge200$ is projected out in the evaluation of physical observables. (a) The $\pi$-Landau-level state with quasienergy $\varepsilon=0.500\hbar\Omega$. (b) Nearly degenerate unphysical state with $\varepsilon=0.499\hbar\Omega$.
(c), (d) Frequency distribution, $\sum_{\sigma=\uparrow,\downarrow}\sum_{nm} |\langle n,\sigma|u_{\alpha,m}\rangle|^2 \delta(\hbar\omega-\varepsilon_\alpha-m\hbar\Omega)$, for the Floquet states shown in (a) and (b), respectively. }
\label{fig:truncation}
\end{figure}

In this appendix, we provide several comments on the details of the numerical calculation. 
As we mention in Sec.~\ref{sec:numerics}, we employ the harmonic oscillator basis to express the Hamiltonian for the linearized Hamilotnian Eq.~(\ref{eq:linearlized-model2}).
Throughout the numerical calculations in this paper (for the linearized Hamiltonian), we truncate the number of harmonic-oscillator modes at $N_b=400$. We have to be careful on the artifact due to this cutoff, because it leads to the unphysical vacuum of $\hat{b}^\dagger$ as a spurious low-energy state, although it has vanishing hybridization with the genuine low-energy states. We eliminate the contribution of $N_b\ge200$ states after the diagonalization. We note that this post process is not necessary for the model on the honeycomb lattice Eq.~(\ref{eq:lattice-model}) since it does not rely on the harmonic-oscillator basis.

To check the validity of the truncation scheme, we plot the distribution of the harmonic-oscillator modes for typical Floquet states in Figs.~\ref{fig:truncation}~(a) and (b). Here we set $B=\hbar /el_0^2$, $E=0$ and $p_y=0.2\hbar\Omega/v_F$. We calculate the amplitude by $\sum_{\sigma=\uparrow,\downarrow}\sum_m|\langle n,\sigma|u_{\alpha,m}\rangle|^2$, with $|n,\uparrow\rangle=(|n\rangle,0)^{\text{T}}$, $|n,\downarrow\rangle=(0,|n\rangle)^{\text{T}}$.
Panel (a) shows the distribution for the $\pi$-Landau level state with $\varepsilon=0.500\hbar\Omega$, which confirms the sufficiency of the cutoff parameter $N_b=400$. Panel (b) is for a nearly-degenerate state with $\varepsilon=0.499\hbar\Omega$, 
which should be excluded since it stems from the unphysical vacuum of $\hat{b}^\dagger$ as can be seen. We can see that the truncation of the amplitude $N_b\ge200$ is sufficient for this purpose.

For the computation of the Floquet state, we also need to truncate the photon number $m$ for $|u_{\alpha,m}\rangle$.
Since the direct computation with diagonalization of the Sambe-space Hamiltonian Eq.~(\ref{eq:sambe}) scales cubically with the truncation, we instead compute the time-evolution operator $U(t,0)$ numerically and construct the Floquet state based on the eigenstate of $U(T,0)$. In this scheme the cutoff for the energy is given by $\sim\pi/\Delta t$ with the time step $\Delta t$, and is advantageous since the computational time scales linearly with $1/\Delta t$. We show the frequency distribution of the Floquet state, $\sum_{\sigma=\uparrow,\downarrow}\sum_{nm}|\langle n,\sigma|u_{\alpha,m}\rangle|^2 \delta(\hbar\omega-\varepsilon_\alpha-m\hbar\Omega)$ in Figs.~\ref{fig:truncation}~(c) and (d), for the Floquet states employed in Panel (a) and (b), respectively.

\bibliographystyle{apsrev4-1}
\bibliography{references}

\end{document}